\documentclass[twocolumn]{aastex631}

\newcommand{\conm}[2]{\langle #1\! \mid\! #2 \rangle}

\usepackage{enumitem}
\usepackage{hyperref}
\usepackage{amsfonts}
\usepackage{amsmath}
\usepackage{amssymb}

\usepackage{cleveref}
\usepackage{multirow}
\usepackage{pgfplots}
\usepackage{placeins}
\usepackage{tikz}
\usepackage{upgreek}
\usepackage{booktabs}
\usepackage{physics}
\usepackage{pifont}

\usepackage{cancel}
\usepackage{graphicx}
\usepackage{color}
\usepackage{xcolor}
\usepackage{xparse}

\NewDocumentCommand{\adi}{O{blue} m}{\textcolor{#1}{#2}}
\def\F{frw}
\def\I{inv}
\def\etaM{{\eta^{\!\text{\tiny \I}}}}
\def\Meta{{M^{\!\text{\tiny \F}}}}
\newcommand{\muFTF}{\mu_d^{\!\text{\tiny \F}}}
\newcommand{\muITF}{\mu_d^{\!\text{\tiny \I}}}
\newcommand{\SFTF}{\mathcal{S}^{\text{\tiny \F}}}
\newcommand{\SITF}{\mathcal{S}^{\text{\tiny \I}}}
\newcommand{\sftf}{\SFTF}

\newcommand{\sigln}{\sigma_{\!\scriptscriptstyle \ln}}
\newcommand{\glnf}[2]{%
G_{\!\scriptscriptstyle \ln}\!\left({#1}\!/{#2}\right)}
\newcommand{\glns}[2]{%
G_{\!\scriptscriptstyle \ln}\!\left({#1\!},{\!#2}\!\right)}
\newcommand{\glnsig}{%
G_{\!\scriptscriptstyle \ln}\!\left(r\!/\!d,\sigln \right)}%,{\!#2}\!\right)}
\newcommand{\gln}{\glnf{r\!}{\!d}}%

\newcommand{\dFTF}{d^{\text{\tiny \F}}}
\newcommand{\dITF}{d^{\text{\tiny \I}}}
\newcommand{\dftf}{\dFTF}
\newcommand{\ditf}{\dITF}
\newcommand{\nftf}{f^{\text{\tiny \F}}}
\newcommand{\nitf}{f^{\text{\tiny \I}}}
\newcommand{\rt}{r^\textrm{tru}}
\newcommand{\vt}{V^\textrm{tru}}
\newcommand{\rh}{\hat{\pmb{r}}}
\newcommand{\nh}{\hat{\pmb{n}}}

\def\ves{V^\textrm{obs}}

\def\vobs{V^\textrm{obs}}
\def\vtru{V}

\def\vtr{\vtru}

\def\vwf{\widetilde V}
\def\vr{\pmb{r}}

\usepackage[T1]{fontenc}

\DeclareRobustCommand{\VAN}[3]{#2}
\let\VANthebibliography\thebibliography
\def\thebibliography{\DeclareRobustCommand{\VAN}[3]{##3}\VANthebibliography}

\newcommand{\be}{\begin{equation}}
\newcommand{\ee}{\end{equation}}
\def\lsim{\mathrel{\rlap{\lower4pt\hbox{\hskip0.5pt$\sim$}}
    \raise1pt\hbox{$<$}}}         %less than or approx. symbol
\def\gsim{\mathrel{\rlap{\lower4pt\hbox{\hskip0.5pt$\sim$}}
    \raise1pt\hbox{$>$}}}         %greater than or approx. symbol

\def\ln{{\rm ln}}

\def\mathbi#1{\textbf{\em #1}}

\def\lsim{~\rlap{$<$}{\lower 1.0ex\hbox{$\sim$}}}
\def\bsim{~\rlap{$>$}{\lower 1.0ex\hbox{$\sim$}}}
\def\kms{\ {\rm km\,s^{-1}}}

\def\hmpc{\ {\rm {\it h}^{-1}Mpc}}

\def\nh{\hat{n}}
\def\rh{\hat{r}}

\def\vr{\mathbi{r}}

\def\apjs{{Astrophys.~ J.~ Suppl.~}}
\def\apjl{{Astrophys.~ J.~ Lett.~}}

\shorttitle{distance and velocity Malmquist bias}
\shortauthors{A. Nusser}

\begin{document}

\title{On distance and velocity estimation  in cosmology}

\author[0000-0002-8272-4779]{Adi Nusser}
\email{adin@technion.ac.il}  

%\altaffiliation{Physics Department}
\affiliation{The Technion Department of Physics, The Technion – Israel Institute of Technology, Haifa 3200003, Israel}

% \author[orcid=0000-0000-0000-0002,gname=Bosque, sname='Sur America']{Forrest Sur Am\'{e}rica} 
% \altaffiliation{Las Campanas Observatory}
% \affiliation{Universidad de Chile, Department of Astronomy}
% \email{fakeemail2@google.com}

\begin{abstract}

Scatter in distance indicators introduces two conceptually distinct systematic biases when reconstructing peculiar velocity fields from redshifts and distances. The first is distance Malmquist bias (dMB) that affects individual distance estimates and can in principle be approximately corrected. The second is velocity Malmquist bias (vMB) that arises when constructing continuous velocity fields from scattered distance measurements: random scatter places galaxies at noisy spatial positions, introducing spurious velocity gradients that persist even when distances are corrected for dMB.
Considering  the Tully–Fisher relation as a concrete example, both inverse and forward formulations yield unbiased individual peculiar velocities for galaxies with the same true distance (the forward relation requires a selection-dependent correction), but neither eliminates vMB when galaxies are placed at their inferred distances. We develop a modified Wiener filter that properly encodes correlations between directly observed distance $d$ and true distance $r$
 through the conditional probability $P(r|d)$, accounting for the 
 distribution of true distances sampled by galaxies at observed distance 
 $d$. Nonetheless, this modified filter yields suppressed amplitude 
 estimates. Since machine learning autoencoders converge to the Wiener 
 filter for Gaussian fields, they are unlikely to significantly improve 
 velocity field estimation. We therefore argue that optimal reconstruction 
 places galaxies at their observed redshifts rather than inferred distances; an approach effective when distance errors exceed $\sigma_v/H_0$, a condition satisfied for most galaxies in typical surveys beyond the nearby volume.

\end{abstract}

\keywords{galaxies: distances and redshifts --- cosmology: observations}

\section{Introduction} 

Peculiar velocities of galaxies, their motions relative to the isotropic  Hubble expansion, provide a direct probe of the matter 
distribution and gravitational dynamics of the Universe. On large scales ($\gtrsim 10$ Mpc), where density fluctuations remain linear or mildly nonlinear, peculiar velocities 
trace the growth of cosmic structure and offer constraints on fundamental cosmological 
parameters, particularly the growth rate $f \equiv d\ln D/d\ln a$, where $D(a)$ is the linear growth factor \citep{Peeb80}. This growth rate is a key discriminator between cosmological models and a sensitive test of general relativity on cosmological scales.

An important  test of gravitational instability theory compares observed peculiar velocities with predictions from the large-scale matter distribution. By computing the gravitational force field from galaxy redshift surveys via the Poisson equation, one can predict the 
expected velocity field and test whether galaxies move as gravitational instability predicts  \citep{NBDL,Nusser2020,LilowNusser2021,carrick15}. Such comparisons have provided strong support for standard $\Lambda$CDM cosmology and yielded robust measurements of the growth rate $f$.

Recovering this field from observations is a formidable task fraught with pitfalls. The basic  framework appears   simple: the difference between observed galaxy redshifts and distance estimates yields line-of-sight peculiar velocities, which are then interpolated onto a spatial grid to construct the three-dimensional velocity field. 

The devil, however, is in the details. While accurate redshifts are straightforward to obtain, especially in the local universe where spectroscopic surveys abound, reliable distance measurements represent a far more difficult task. Distances must be inferred from empirical correlations between intrinsic and observable galaxy properties, relations that are inherently noisy. In this work we focus on the Tully-Fisher relation (TF)  \citep{TF77} relation for spiral galaxies, which correlates luminosity with rotation velocity. The difficulty lies not merely in the existence of intrinsic scatter in these relations, but in the practical challenge of acquiring accurate measurements of the intrinsic properties themselves, which for TF  means obtaining high-quality  line widths, a demanding observational task that limits sample sizes and introduces selection effects. Consequently, distance measurements are sparse and distributed non-uniformly across the sky, rendering the task of inferring a continuous three-dimensional velocity field from these scattered data points highly non-trivial. This is further complicated  by the fact that observations constrain only the line-of-sight component of the peculiar velocity vector, leaving the transverse components entirely undetermined by the data. Although our discussion centers on TF, the conceptual framework and systematic biases we identify apply equally to other distance indicators such as the Fundamental Plane for elliptical galaxies \citep[e.g.,][]{Fundamental_Plan87,Saulder_Fundamental,CF4}.

% Observationally, peculiar velocities are inferred by combining redshift measurements with independent distance estimates. For spiral galaxies, the Tully-Fisher (TF) relation between luminosity and rotational velocity \citep{TF77} serves as the primary distance indicator. Given a measured redshift $cz$ and an inferred distance $d$, the line-of-sight peculiar velocity is simply $v = cz - H_0 d$. However, this apparently straightforward procedure is fraught with systematic biases that can severely compromise large-scale velocity field reconstructions.

One class of bias arises from sample selection, particularly magnitude limits that preferentially exclude faint galaxies at large distances. Another, more subtle bias stems from the fact that distance estimates scattered by measurement errors are biased estimators of true distances, especially in the presence of spatial density gradients \citep{Lyn88}. Following \citet{SW95}, we refer to the first as \emph{selection bias} and the second as \emph{Malmquist bias}, though both trace their origins to Malmquist's pioneering work in the 1920s. The latter is especially difficult to correct and has remained a major obstacle to reliable velocity field reconstruction on scales of a few Mpc.

In this paper, we make a fundamental distinction between two types of Malmquist bias that have not been clearly separated in the 
literature: \emph{distance} Malmquist bias (dMB), which affects individual distance estimates, and \emph{velocity} Malmquist bias (vMB), which arises when constructing continuous velocity fields 
from these scattered estimates. We demonstrate that while dMB can, in principle, be approximately corrected using methods such as the Feast-Landy-Szalay (FLS) prescription \citep{Feast1972,Landy1992}, vMB is a more fundamental limitation that cannot be removed by correcting 
individual distance measurements. The key insight is that even when distance estimates are unbiased on average, the random scatter in these estimates introduces systematic biases into the reconstructed 
velocity field that persist regardless of corrections applied to individual galaxies.

We focus on the TF relation as an example, though our conclusions apply to any distance indicator based on intrinsic 
scaling relations with significant scatter. There are two equivalent formulations: the \emph{forward} TF relation predicts absolute 
magnitude from observed rotational velocity, while the \emph{inverse} TF relation predicts velocity from absolute 
magnitude. Because of intrinsic scatter, these two relations are not mathematically consistent---the inverse of a scattered linear 
relation is not the same relation inverted. The inverse formulation has a practical advantage: observational selections on velocity are typically weaker than on magnitude, making selection effects more tractable.

However, we show that obtaining unbiased distances does not guarantee an unbiased velocity field. The random scatter that 
remains even after bias corrections causes galaxies to be assigned to incorrect spatial positions, introducing spurious velocity 
gradients and correlations. We demonstrate through both analytical arguments and numerical simulations that the optimal strategy for velocity field reconstruction is not to correct individual distances, but rather to assign galaxies to their observed 
\emph{redshift} coordinates, which serve as proxies for true distances. This approach minimizes vMB on scales large compared to the velocity dispersion ($\gtrsim 3\,h^{-1}$ Mpc) and provides a more robust foundation for cosmological analyses.

At this stage  it is useful to sharpen  what we mean by a "biased" field. Consider a straightforward 1D top-hat smoothing of a zero-mean field $f(x)$. Denote  the smoothed field by $f\textrm{S}$, then   $\conm{f(x)}{f^\textrm{S}}=\langle f f^\textrm{S}\rangle/\langle f ^\textrm{S}f^\textrm{S}\rangle f^\textrm{S}\ne f^\textrm{S} $. Therefore, $f^\textrm{S}$ is a biased representation of $f$. However, we consider this a trivial bias since it depends solely on the properties of the field $f$. When we speak of vMB, we have in mind a bias that depends on an unknown underlying density field, namely the underlying number density $n(r)$ of objects, which as we shall see appears in expressions for the conditional PDFs of $P(r|d)$. 

The paper is organized as follows. 
Section~\ref{sec:prelim} establishes our notation and reviews the forward and inverse Tully--Fisher relations, highlighting their self-consistency issues in the presence of scatter. 
Section~\ref{sec:distance_inference} derives distance estimators from the forward and inverse relations, and Section~\ref{sec:dMB} analyzes distance Malmquist bias and its correction, including the FLS approach. 
Section~\ref{sec:velfield} introduces velocity Malmquist bias and examines several strategies for peculiar–velocity and velocity–field reconstruction. 
Section~\ref{sec:toy_model} illustrates these effects using a spherically symmetric toy model, and Section~\ref{sec:ml_wiener} discusses machine–learning approaches and a modified Wiener filter for mitigating vMB, supported by a particle–mesh simulation. 
Appendix~\ref{sec:fundamental} extends our formalism to the Fundamental Plane, and Appendix~\ref{sec:graziani} assesses recent Bayesian methods based on Gibbs sampling. 
We summarize our results and discuss their implications in Section~\ref{sec:discussion}.

\begin{table*}[ht!]
\centering
\caption{Comparison of forward and inverse Tully--Fisher formulations. 
The log-Gaussian kernel is $G_{\ln}(r/d) \equiv \exp[-(\ln(r/d))^2/(2\sigma_{\ln}^2)]$. 
See \S\ref{sec:viaFTF}--\ref{sec:LS} for derivations.}
\renewcommand{\arraystretch}{1.5}
%\begin{tabular}{p{0.28\linewidth} p{0.32\linewidth} p{0.32\linewidth}}
\begin{tabular}{lll}
\toprule
 & \textbf{Forward TF (fTF)} & \textbf{Inverse TF (iTF)} \\
\midrule
Relation 
& $M = a\eta + b + \epsilon_M$ 
& $\eta = \gamma M + \eta_0 + \epsilon_\eta$ \\
\midrule
Distance modulus estimator
& $\muFTF = m - a\eta - b$ 
& $\muITF = \gamma^{-1}(\eta_0 - \eta) + m$ \\
\midrule
Is $\langle \ln d | \ln r \rangle = \ln r$\,? 
& Yes if $\SFTF(d) = 1$ 
& Yes \\
\midrule
Selection enters $P(d|r)$? 
& Yes, via $\SFTF(d)$ 
& No \\
\midrule
Selection enters $P(r|d)$? 
& No 
& Yes, via $\SITF(r)$ \\
\midrule
%Bias correction 
%& Estimate $\SFTF(d)$ from data (\S\ref{sec:SFTF}); apply before velocity estimation
%& Multiply $d$ by $e^{-\sigma_{\ln}^2/2}$; \\ no selection correction needed for $P(d|r)$ \\
%\midrule
%Advantage
%& Selection function $\SFTF$ can be measured directly from magnitude-limited catalogs
%& Distance modulus unbiased by construction; simpler error propagation \\
\bottomrule
\end{tabular}
\label{tab:tf_comparison}
\end{table*}

%%%%%%%%%%%%%%%%%%%%%%%%% fTF and iTF  %%%%%%%%%%%%%%%%%%%%%%%%%%%%%%%
\section{Forward and Inverse relations and their self-consistency}
\label{sec:prelim}
The distance modulus, $\mu$, of an object with comoving coordinate $r$ and
luminosity distance $d_L(r)$ is
\begin{equation}
\label{eq:DM}
\mu_r = m - M
    = 5\log_{10}\!\left(\frac{d_L(r)}{10\,\mathrm{pc}}\right)
    = \frac{5}{\ln 10}\,\ln d_L(r) + \mathrm{const}\, ,
\end{equation}
where $m$ and $M$ are the apparent and absolute magnitudes of the object.

For simplicity, we restrict the equations to the Tully–Fisher (hereafter TF)
relation between the luminosities $L$ and rotational velocities
$v_\textrm{rot}$ of spiral galaxies.  
As shown in Appendix~\ref{sec:fundamental}, the same results apply to the
Fundamental Plane distance indicator, which involves three galaxy parameters
rather than two as in the TF relation.

To streamline notation, we define the log–Gaussian kernel
\begin{equation}
\gln \equiv 
\frac{1}{\sqrt{2\pi}\,\sigma_{\ln}}
\exp\!\left[-\frac{(\ln(r/d))^{2}}{2\sigma_{\ln}^{2}}\right],
\end{equation}
where $\sigma_{\ln}^{2}$ is the variance of $\ln(r/d)$.

We distinguish between the \textit{forward} Tully–Fisher (fTF) and the
\textit{inverse} TF (iTF) formulations.  
Neglecting observational selections, the fTF relation expresses the
\emph{absolute magnitude} $M$ at fixed \emph{linewidth parameter}
$\eta = 2\log v_\textrm{rot}$ as
\begin{equation}
M = M_{\rm frw}(\eta) + \epsilon_M ,
\label{eq:FTF_raw}
\end{equation}
where
\begin{equation}
M_{\rm frw}(\eta) = a\,\eta + b ,
\label{eq:FTF}
\end{equation}
and $\epsilon_M \sim \mathcal{N}(0,\sigma_M^2)$ represents the intrinsic
scatter around the forward TF relation.

In the iTF formulation, the distance-dependent magnitude $M$ is instead used to
predict the linewidth parameter:
\begin{equation}
\eta = \eta_{\rm inv}(M) + \epsilon_\eta ,
\label{eq:ITF_raw}
\end{equation}
where
\begin{equation}
\eta_{\rm inv}(M) = \gamma\,M + \eta_0 ,
\label{eq:ITF}
\end{equation}
and $\epsilon_\eta \sim \mathcal{N}(0,\sigma_\eta^2)$ represents the intrinsic
scatter.  
Because of this scatter, the slopes in the two formulations are not simple
inverses, and in general $\gamma \neq 1/a$.
 
 \subsection{Self-consistency}
The conditional PDF of $M$ given $\eta$ is  
\begin{equation}
P(M \mid \eta)
  \;=\;
  \frac{1}{\sqrt{2\pi}\,\sigma_M}
    \exp\!\left[-\frac{\big(M - \Meta(\eta)\big)^2}{2\sigma_M^2}\right] \; ,
    \label{eq:pdfFTF}
\end{equation}
and similarly for $P(\eta\mid M)$. 
Bayes' theorem links the two relations,
\begin{equation}
P(\eta \mid M) \;=\; \frac{P(M \mid \eta)\,P(\eta)}{P(M)} \; .
\end{equation}

Using \cref{eq:pdfFTF}, the last equation can be written as 
\begin{equation}
P(\eta \mid M) \;\propto\;
\exp\!\left[-\frac{\big(M - \Meta(\eta)\big)^2}{2\sigma_M^2}\right]\,
P(\eta)\,P(M)^{-1} \; .
\end{equation}

If both $P(M\mid\eta)$ and $P(\eta\mid M)$ are Gaussian, the joint distribution
\(P(M,\eta)\) must be \emph{bivariate normal}, implying Gaussian marginals for
both $P(M)$ and $P(\eta)$.  However, in reality the luminosity distribution of
spirals follows a Schechter function rather than a Gaussian, so $P(M)$ is
strongly non-Gaussian.  Consequently, $P(M\mid\eta)$ and $P(\eta\mid M)$ \emph{cannot
both be exactly Gaussian}.  The usual practice of adopting Gaussian scatter in
either the forward or the inverse TF relation therefore provides only an
\emph{approximate} statistical description.  The two formulations are not
mutually consistent at a fundamental probabilistic level, although the
difference is often negligible when the intrinsic scatter is small and the
dynamic range in $M$ or $\eta$ is limited.

% JOINT PDF 

\section{Distance inference}
\label{sec:distance_inference}
Most of the basic relations here are based on the the paper of \cite{SW95} (herafter SW95). 
In the following, we define observed distances that can be directly derived from either the forward or inverse TF relations. These definitions rely solely on the quantities entering the distance indicator itself, which for the TF relations are  the magnitudes and linewidths, and require no additional information. We then discuss the properties of these distances and revisit the question of how they are biased with respect to the true distances. Table~\ref{tab:tf_comparison} summarizes the key statistical
properties of the main distance estimators discussed  in this section.

\subsection{Distance via the  fTF relation}

We define an observed (inferred) distance $d$ in terms of the distance modulus 
$\mu_d=\mu(d)$, as 
\begin{equation}
 \mu_d= m-\Meta(\eta)=m-a \eta -b \; .
\label{eq:dFTF}
\end{equation}
This obviously differs  from the true distance modulus, $\mu_r$, defined in \cref{eq:DM}, such that 
\begin{equation}
    \mu_d = \mu_r +\epsilon_M \; .
\end{equation}
In the ideal case of the absence of observational selections, this last relation  implicitly implies that the mean of all possible $\mu_d $ corresponding to  the same true $\mu_r $ is equal to $\mu_r$.
It does not, however,  specify on its own  $\mu_r$ for a the same  observed $\mu_d$.

In order to determine the general statistical relation between $\mu_r$ and $\mu_d$ we follow SW95 and resort to the joint  probability density, $P(r,m,\eta)$, for true distance $r$, apparent magnitude $m$,
and the width parameter $\eta$.  
Assuming the fTF relation \cref{eq:FTF}, we write 
\begin{align}
& P(r,m,\eta)= P(M=m-\mu_r,\eta\mid r)P(r) \notag\\[4pt]
&\propto\; r^{2}\,n(r)\,S(m,\eta)\,\phi(\eta)\,
\,
\exp\!\left[-\frac{\bigl(m -\mu_r - \Meta(\eta)\bigr)^{2}}{2\sigma_M^{2}}\right].
\end{align}
where $n(r)$ is the real–space density, $S(m,\eta)$ the selection function,
$\phi(\eta)$ the distribution of  $\eta$, and $\sigma$ the TF scatter.

The joint $PDF$, $P(r,\muFTF)$ is obtained  from 
\begin{equation}
P(r,d)=\int \dd m \dd \eta P(r,m,\eta)\delta^\textrm{D}\!\left(\mu_d -m+\Meta(\eta)\right) \; , 
\end{equation}
where $\delta^\textrm{D}$ is the Dirac-delta function. 
Performing the integration, we arrive at 
\begin{equation}
P(r,d) \propto r^2 n(r)\exp\!\left[-\frac{\bigl(\mu_d-\mu_r\bigr)^{2}}{2\sigma_M^{2}}\right]\frac{\SFTF(d)}{d}\; ,
\label{eq:Prdftf}
\end{equation}
where 
\begin{equation}
\SFTF(d)=\int \dd \eta \phi(\eta) S\left(m=\mu_d+\Meta(\eta),\eta\right)\; .
\label{eq:SFTF}
\end{equation}

Thus the PDF of $d$ given $r$ is 
% \begin{equation}
% \label{eq:Pdrcondftf}
% P(d\mid r)=\frac{\SFTF(d) d^{-1}\exp\!\left[-\frac{\bigl(\ln(r/d)\bigr)^{2}}{2\sigma_{\ln}^{2}}\right]}{\int \dd d\,  \SFTF(d)d^{-1}\exp\!\left[-\frac{\bigl(\ln(r/d)\bigr)^{2}}{2\sigma_{\ln}^{2}}\right]}\; ,
% \end{equation}
\begin{equation}
\label{eq:Pdrcondftf}
P(d\mid r)=\frac{\SFTF(d) d^{-1}\, \glnsig }{\int \dd d\,  \SFTF(d)d^{-1}\, \glnsig}\; ,
\end{equation}
where we  $\mu_d-\mu_r=(5/\ln 10)\ln(r/d)$ and  $\sigma_{\ln} = (\ln 10/5) \sigma_M$. Conditioning on $d$ gives the normalized conditional PDF (posterior),
\begin{equation}
\label{eq:Prdcondftf}
P(r\mid d)
=\frac{r^2 n(r) \glnsig} {\int \dd r r^2 n(r) \glnsig}\; ,
\end{equation}
% \begin{equation}
% \label{eq:Prdcondftf}
% P(r\mid d)
% =\frac{r^2 n(r) \exp\left[-\frac{(\ln (r/d))^2}{2\sigma^2}\right]}{\int \dd r r^2 n(r) \exp\left[-\frac{(\ln (r/d))^2}{2\sigma^2}\right]}\; ,
% \end{equation}
which,  unlike $P(d\mid r)$, is independent of the selection imposed on the survey and depends on the underlying galaxy density $n(r)$ from which the catalog galaxies were selected. 

In  Appendix~\ref{sec:SFTF}, we present a method for  determining $\SFTF$ directly from the data, similar to the algorithm of \citet{dav82} developed for evaluating selection functions corresponding to magnitude-limited redshift surveys.

\label{sec:viaFTF}

\subsection{Distance Via the iTF relation}
We now derive the analogous expressions for the inverse TF.
Given measured $\eta$ and $m$, the distance is  
inferred by setting $\etaM(M) = \eta$, i.e.,
\begin{equation}
\muITF = \gamma^{-1} (\eta_0 - \eta) + m \; .
\label{eq:dITF}
\end{equation}

In this case, 
\begin{equation}
P(r,m,\eta)\propto r^{2} n(r) S(m,\eta) \Phi(M)
\exp\left[-\frac{\bigl(\eta - \etaM(M)\bigr)^{2}}{2\sigma_\eta^{2}}\right]\; ,
\end{equation}
where $M = m - \mu_r$ and $\Phi(M)$ is the galaxy luminosity function. 

Following similar steps to those in \S\ref{sec:viaFTF}, we obtain
\begin{equation}
P(r,d) \propto r^2 n(r) \SITF(r,d) \glnsig \; ,
%\exp\left[-\frac{(\ln(r/d))^2}{2\sigma^2}\right]\; ,
\label{eq:Prditf}
\end{equation}
where now  $\sigln =(5/\ln 10) \sigma_\eta / \gamma$, and,
\begin{equation}
\SITF(r,d) = \int \dd m\, \Phi\left(m - \mu(r)\right)
S\left(m, \etaM(m - \mu_d)\right)\; .
\end{equation}
In SW95, the function $\SFTF(r,d)$ is assumed to depend only on $r$. 
This is strictly correct if $S(m,\eta)$ does not depend explicitly on $\eta$. 
However, they argue that $\SITF(r,d)$ effectively becomes a function of $r$ if the dependence on $r$ is weak, allowing $\mu_d$ to be replaced by $\mu_r$. 
Here, we will assume that explicit selection on $\eta$ is negligible compared to that on $m$, and hence take $\SITF(r,d) = \SITF(r)$.
Therefore, the conditional PDFs become 
\begin{equation}    
P(d|r)= \frac{d^{-1} \glns{r/d}{\sigma_{\ln}}}{\int \dd d\, d^{-1}\glns{r/d}{\sigma_{\ln}}} \; ,
    \label{eq:Pdrconditf}
\end{equation}
and
\begin{equation}
    P(r|d) = \frac{r^2 n(r) \SITF(r) \glnsig}{\int \dd r r^2 n(r) \SITF(r) \glnsig}\; .
    \label{eq:Prdconditf}
\end{equation}

\subsection{Origin of difference between iTF and fTF distances}

The distance moduli $\muFTF$ and $\muITF$, derived respectively via \cref{eq:dFTF} and \cref{eq:dITF}, applied to the same data set, differ by more than a constant only because $\gamma^{-1}\ne a$.
The residual between the two distance moduli for the same object is
\begin{equation}
\muFTF-\muITF = (\gamma^{-1}-a)\eta + const; .
\end{equation}
Since $\conm{\muITF}{r}=\mu_r$, this implies that any non-constant difference between the two estimates originates from a non-trivial dependence of $\conm{\eta}{r}$ on the true distance $r$.
This dependence arises from the selection on $m$. To see this, imagine a scatter plot of $M$ versus $\eta$ for galaxies at the same distance. The cut in $m$ removes all objects fainter than a certain $M$, thereby altering the mean $\eta$ of the remaining galaxies. Because a constant threshold in $m$ corresponds to a distance-dependent cut in $M$, the mean $\eta$ acquires a distance dependence.

\section{Distance Malmquist Bias }
\label{sec:dMB}
We are interested in estimating the true distance given an observed distance $d$.  
Before addressing this main question, it is important  to first examine the properties of distances $d$ inferred from measurements of objects with the  true distances close to   $r$. This is relevant for the estimation of galaxy peculiar velocities discussed in \S\ref{sec:velfield}.

The form in \cref{eq:Pdrcondftf} for $P(d\mid r)$ in the fTF relation implies, for a general $\SFTF(d)$, that  
\begin{equation}
\conm{\ln \dFTF}{r} = \int \dd d\, P(d\mid r)\,\ln d \ne \ln r \; .
\end{equation}
For the \textit{ideal case only} of $\SFTF=1$ in \cref{eq:Pdrcondftf},
\begin{equation}
\conm{\ln \dFTF}{r} = \ln r  \qq{and} \conm{\dftf}{r} = r\exp\!\left(\frac{\sigma_\mathrm{ln}^2}{2}\right)\; ,
\end{equation}
where the trivial bias in the second equality arises from the lognormal nature of the distribution and can be easily corrected.
When selection effects in $m$ and $\eta$ are present, however, the correction requires knowledge of $\SFTF$. Fortunately, as we shall see below, $\SFTF$ can be inferred directly from the data if explicit selection on $\eta$ can be neglected. 

Using \cref{eq:Pdrconditf}, the iTF satisfies, for any $\SITF(r)$,
\begin{equation}
\conm{\ln \dITF}{r} = \ln r \qq{and} \conm{\ditf}{r} = r\exp\!\left(\frac{\sigln^2}{2}\right)\; ,
\label{eq:EITF}
\end{equation}
which, unlike the fTF case, holds even when selection is applied, provided that the explicit dependence of the selection on $\eta$ is weak.

Having described the behavior of $d$ for a given $r$, we now turn to the  problem of determining how the true distance $r$ relates statistically to an observed $d$. 

For the fTF, from \cref{eq:Prdconditf} it is straightforward to see that
\begin{equation}
\conm{\ln r}{\dFTF} \ne d \qq{and}  \conm{r}{\dFTF} \ne \dFTF \; .
\end{equation}
This bias in the conditional mean of the true distance is known as the \textit{spatial} Malmquist bias \citep{Lyn88}. 
Here we refer to it as the \textit{distance} Malmquist bias (dMB) to distinguish it from the \textit{velocity} Malmquist bias (vMB), discussed later.

For a homogeneous galaxy distribution, $n(r)=\mathrm{const}$, one obtains for the fTF case
\begin{equation}
\conm{\ln r}{\dftf} = \ln \dftf + 3\sigln^2 \qq{and} 
\conm{r}{\dftf} = \dftf \exp\!\left(\frac{7\sigln^2}{2}\right)\; .
\label{eq:rfTF}
\end{equation}
Similarly, with  $n(r)=const$ for the iTF  the selection function $\SITF(r)$ enters explicitly, giving
\begin{equation}
\conm{\ln r}{\ditf} = \ln \ditf + 3\sigln^2 + 
\frac{\dd \ln \SITF}{\dd \ln r}\,\sigln^2
\end{equation}
and 
\begin{equation}
\conm{r}{\ditf} = \ditf \exp\!\left[\frac{7\sigln^2}{2} + 
\frac{\dd \ln \SITF}{\dd \ln r}\,\sigln^2\right]\; ,
\label{eq:riTF}
\end{equation}
where the additional term proportional to $\dd \ln \SITF/\dd \ln r$ accounts for the effect of distance-dependent selection in the iTF case.

The focus here is not on the trivial bias arising from the lognormal nature of the distance modulus, but on the spatial component of the Malmquist bias. 
Recovering unbiased estimates of true distances requires accounting for the underlying galaxy distribution and any selection effects. 
In the fTF case, this involves the density field entering \cref{eq:Prdcondftf}, while for the iTF it additionally depends on the selection function $\SITF(r)$ appearing in \cref{eq:Prdconditf}. 
Consequently, the conditional mean $\conm{r}{d}$ depends non-trivially on the observed $d$ through factors such as $r^{2}n(r)$ in the fTF and $r^{2}n(r)\SITF(r)$ in the iTF.

\subsection{dMB correction:  generalizing Feast and Landy \& Szalay to both iTF and fTF}
\label{sec:LS}

\cite{Feast1972} and \cite{Landy1992} (FLS) offered an analytic formula for 
computing $\conm{r}{d}$ from the distribution of  observed distance alone. 
This method is applicable for the iTF analysis only. 
We define  $\nitf(d)\dd d$ as the number of objects with observed iTF distances between $d$ and $d+\dd d$, per solid angle.
Using $\nitf(d)  = \int \dd r P(r, d)$ with the posterior $P(r,d) $ given in \cref{eq:Prditf}, it can be shown that,
\begin{equation}
\label{eq:Erdls}
\bar r(d)=\conm{r}{d} =d \mathrm{e}^{3\sigma_\mathrm{ln}^2/2}\frac{\nitf(d\mathrm{e}^{\sigma_
\mathrm{ln}^2})}{\nitf(d)} \; ,
\end{equation}
The  r.h.s of this expression depends solely on the distribution of galaxies in the observed distance space. However, estimating the expression of the data is not trivial as the ratio of the density $f$ at two different distances can be very noisy.

The same trick cannot readily be used for the fTF since according to \cref{eq:Prdcondftf}  
the posterior $P(r|d) $ in this case is independent of the function $\sftf(d)$, while $P(r,d)$ in \cref{eq:Prdftf} and hence $\nftf(d)=\int \dd r P(r,d) $ depends on this function. However, 
it is easy to see that using 
\begin{equation}
    g(d)\equiv \frac{\nftf(d)}{\sftf(d)}
\end{equation} instead of $\nitf$ in  \cref{eq:Erdls} yields 
 $\conm{r}{d}$ in the fTF case. Since we have shown in \S\ref{sec:SFTF}
that $\sftf(d)$ can also be derived from the data directly then the FLS method can be extended to the fTF as well.

\section{Peculiar Velocity Estimation and velocity Malmquist bias}
\label{sec:velfield}

Distances are of interest in their own right, especially when they can be estimated to sufficiently large redshifts such that the distance--redshift relation constrains the matter and dark energy content of the Universe. Here, however, we are primarily concerned with peculiar velocities that can be inferred from distance indicators combined with galaxy redshifts.

There are two related aspects to peculiar velocity estimation: the determination of peculiar velocities for individual galaxies, and the reconstruction of the velocity field on a spatial grid. In both cases one faces the fundamental problem of assigning galaxies to their proper spatial positions, since the true distances are not directly known.

We continue to denote the true comoving distance to a galaxy by $r$, while the observed distance inferred from either the forward or inverse TF relation is written as $d$. When applying the Landy--Szalay posterior method, we distinguish between a random sample from the posterior $P(r|d)$, denoted $\rh$, and the posterior mean distance $\bar{r} = \langle r|d\rangle$. The redshift-space coordinate is written as $s = cz/H_0$.

The key distinction throughout this analysis is between two separate choices: first, the velocity \emph{estimate} assigned to a galaxy (such as $s-d$, $s-r$, or some corrected variant), and second, the spatial \emph{coordinate} at which that velocity is placed when constructing a field (which could be $r$, $d$, $\rh$, $\bar{r}$, or $s$). As we shall demonstrate, even unbiased individual velocity estimates can produce biased velocity fields when galaxies are assigned to scattered or biased distance coordinates.

Observed redshifts are required in order to derive estimates for the line-of-sight peculiar velocities of objects with measured distances. A galaxy redshift coordinate is written as
\begin{equation}
s = \frac{cz}{H_0} = r + \frac{\vtru_r}{H_0}\, ,
\end{equation}
where $\vtru_r$ is the true peculiar velocity of the galaxy along the line of sight (expressed in units of distance by dividing by $H_0$). Therefore, given galaxy redshifts $s$ and observed distances $d$, the most direct estimate of the line-of-sight peculiar velocity is\footnote{An alternative form, valid when $\vobs \ll s$, relates the velocity to the difference between the apparent and inferred distance moduli, $\mu_s$ and $\mu_d$, respectively, as $\vobs \simeq 5\log e\,(\mu_s-\mu_d)s$ \citep{Nusser1995,Watkins2015}. This follows from expanding $\mu_d = 5\log(s - \vobs) + {\rm const}$ to first order in $\vobs/s$.} 
\begin{equation}
\vobs = s - d\, .
\label{eq:sV}
\end{equation}

In the iTF formulation, once corrected by the factor $\exp(\sigma^{2}/2)$ (see Eq.~\ref{eq:EITF}), the observed distance $\ditf$ is unbiased with respect to the true distance $r$. Therefore,
\begin{equation}
\conm{\vobs}{r} = \conm{s - d}{r} = r + \vtru(r) - r = \vtru(r)\, ,
\end{equation}
demonstrating that the velocity estimate is unbiased when conditioned on the true distance.

In the fTF formulation the estimated distance $\dftf$ is biased due to the presence of the selection function $\SFTF(d)$. We assume here that an $\SFTF$ correction has been properly applied (cf.~\S\ref{sec:SFTF}). In this case, $\conm{\vobs}{r} = \vtru(r)$, where $\vtru(r)$ is the true line-of-sight peculiar velocity. The essential result is that individual galaxy velocity estimates from either the iTF (with lognormal correction) or fTF (with $\SFTF$ correction) are unbiased \emph{when conditioned on the true distance}. The challenge arises when we attempt to construct a continuous velocity field, since we do not have access to the true distances.

We now consider several strategies for constructing an estimate of the velocity field $V(\mathbf{x})$, where $\mathbf{x}$ denotes spatial coordinates. The central difficulty is that, although individual galaxy velocity estimates can be unbiased when conditioned on the true distance, the resulting \emph{field} may nonetheless be biased if galaxies are assigned to scattered or systematically biased coordinates. This \emph{velocity Malmquist bias} (vMB) represents a deeper limitation than distance Malmquist bias alone: even perfectly bias-corrected distance estimates do not guarantee an unbiased velocity field.
Table~\ref{tab:velocity_estimators} summarizes the properties of commonly used velocity–field estimators, highlighting both the velocity assigned to each object and the coordinate at which that velocity is placed.

\begin{table*}[ht!]
\centering
\caption{
Velocity field estimators and their bias properties.
The key distinction is between (i) the velocity estimator assigned to each galaxy
and (ii) the spatial coordinate at which that velocity is placed.
Even unbiased per-galaxy velocities generally produce biased fields
when galaxies are assigned to noisy or model-dependent distance coordinates.
}
\renewcommand{\arraystretch}{1.4}
%\begin{tabular}{llcccp{0.36\textwidth}}
\begin{tabular}{llcccl}
\toprule
Estimator &
Velocity &
Coordinate &
Directly observable? &
Field unbiased? &
Comments \\
\midrule
%\multicolumn{6}{l}{\textit{Ideal reference (not observable in practice)}} \\
%\midrule
$\vtru(r)$ &
$s - r$ &
$r$ &
No &
Yes &
Ground truth; $r$ unknown \\
\midrule
\multicolumn{6}{l}{\textit{Placing galaxies at inferred or model-dependent distance coordinates}} \\
\midrule
$\vobs(d)$ &
$s - d$ &
$d$ &
Yes &
No &
Observed TF distance; strong vMB \\
& & & & & due to scatter in $d$; produces spurious gradients \\ 
& & & & & even if $V = 0$ \\
\midrule
$\bar{V}(\bar{r})$ &
$s - \bar{r}$ &
$\bar{r} = \langle r \mid d \rangle$ &
No &
No &
In principle $\bar r $ can be inferred from data via \\ 
& & & & & FLS. Thus removes dMB but vMB remains \\ 
& & & & & due to scatter in $\bar{r}$ \\
\midrule
$\hat{V}(\hat{r})$ &
$s - \hat{r}$ &
$\hat{r} \sim P(r \mid d)$ &
No &
No &
Posterior samples; require $P(r \mid d)$ and $n(r)$; \\
& & & & &  $\langle r \mid \hat{r} \rangle \neq \hat{r}$; not observationally accessible \\
\midrule
%%%
$\vtru(d)$ &
$s - r$ &
$d$ &
No &
No &
True velocities placed at wrong coordinates; \\
& & & & & illustrates coordinate-induced field bias \\
\midrule
\multicolumn{6}{l}{\textit{Placing galaxies at redshift coordinates (recommended)}} \\
\midrule
$\vobs(s)$ &
$s - d$ &
$s$ &
Yes &
Yes &
Unbiased on scales $\gtrsim \sigma_v / H_0$; residual bias \\ 
& & & & & $\propto \sigma_v^2 / s$ is negligible for $s \gg$ few Mpc \\
\bottomrule
\end{tabular}
\label{tab:velocity_estimators}
\end{table*}

\subsection{Strategy 1: Placing Galaxies at Observed Distances}
\label{sec:velocity_at_d}

Consider the velocity field obtained by placing galaxies at their observed distances $d$. We define $\vobs(d) $ as mean velocity of all objects with  observed  $d$, i.e. $
\vobs(d)=\conm{V}{d}$. Given the conditional PDF, $P(r|d) $ of true distance $r$ given $d$, we have  
\begin{equation}
\label{eq:vobsd_bias}
\begin{aligned}
\vobs(d) &= \int (s-d)\,P(r|d)\,{\rm d} r \\
         &= \int \bigl(\vtru(r)+r-d\bigr)\,P(r|d)\,{\rm d} r \\
         &= \conm{\vtru}{d} + \conm{r}{d} - d\, ,
\end{aligned}
\end{equation}
where we have used $s = r + \vtru(r)$ and defined the conditional expectation
\begin{equation}
\conm{\vtru}{d} = \int \dd r\, \vtru(r)\, P(r|d)\, .
\end{equation}
Note that $\conm{\vtru}{d}\ne Vr=(d)$, i.e., it is not  the true velocity evaluated at $r=d$.

The term $\conm{r}{d} - d$ represents the dMB correction discussed above. However, the conditional expectation $\conm{\vtru}{d}$ is itself a biased representation of the true field. 
To see that we compare  this conditional expectation with  actual velocity field 
evaluated at position $r=d$, i.e., $\vtru(r=d)$.
Since $P(r|d) $ is weighted by $r^2n(r)$, $\conm{\vtru}{d}$ could be mostly governed  by $\vtr$
at a position $r\ne d$ rather than $r=d$, and not merely a smoothed version of the true field.

The velocity field $\vobs(d)$ obtained by placing galaxies at their observed distances $d$ exhibits severe vMB. It leads spurious flow even if the true velocity is $V(r)=0$. This approach should generally be avoided unless specific corrections are applied  as we propose in \S\ref{sec:ml_wiener}.

\subsection{Strategy 2: Placing Galaxies at Bias-Corrected Distances}
\label{sec:velocity_at_rbar}

One might hope that using FLS-corrected distances $\bar{r} = \conm{r}{d}$ would eliminate the vMB. We examine two implementations of this strategy, both of which fail to remove the bias, although it is much more reduced in the second.

The first implementation assigns the observed  velocity  $\ves = s - d$ to each galaxy but places it at the corrected distance $\bar{r}$. The mean velocity at coordinate $\bar{r}$ is then
\begin{equation}
\conm{\ves}{\bar{r}} = \conm{r}{\bar{r}} + \conm{\vtru}{\bar{r}} - \conm{d}{\bar{r}}\, .
\end{equation}
If the mapping $\bar{r}(d)$ is one-to-one (which it generally is not in the presence of generic density fluctuations), then $\conm{\ves}{\bar{r}} = \conm{\ves}{d(\bar{r})}$, which is identical to Eq.~(\ref{eq:vobsd_bias}) for $d = d(\bar{r})$. Therefore, this estimator remains biased and like \textit{strategy I} generates  spurious  flows even when $V(r)=0$.

The second approach defines a velocity estimate $\bar{V} = s - \bar{r}$ and places it at $\bar{r}$. The mean velocity becomes
\begin{equation}
\label{eq:second_impl}
\conm{\bar{V}}{\bar{r}} = \conm{r + \vtru - \bar{r}}{\bar{r}} = \conm{\vtru}{\bar{r}}\, ,
\end{equation}
where we have used $\conm{r}{\bar{r}} = \bar{r}$ by construction. This simplifies to $\conm{\vtru}{d(\bar{r})}$, which again differs from $\vtru(r=\bar{r})$ due to the biased weighting of $P(r|d)$. While this  bias here is less sever than the first implementation as it does not produce  unphysical spurious velocities when the true velocity is $V(r)=0$, it nevertheless fails to recover the true field.

The physical meaning of $\conm{\vtru}{\bar{r}}$ is that it represents the mean true velocity of all galaxies with observed distance $d$ corresponding to $\bar{r}$. This differs from the set of galaxies that lie at $r= \bar{r}$, some of which have observed distances different from $d$. To see this, consider a velocity field that vanishes everywhere except in a narrow shell around $r=r_1$. Galaxies with true positions near $r_1$ can have observed distances $d$ that place them at $\bar{r} \neq r_1$, causing $\conm{\vtru}{\bar{r}}$ to be nonzero even though $\vtru(r=\bar{r})=0$.

Neither implementation removes vMB, because the fundamental problem is the stochastic scatter in $d$ around $r$, not merely the systematic offset. The posterior $P(r|\bar{r})$ still reflects the uncertainty in the true distance given the observed $d$, and this uncertainty propagates into systematic biases in the velocity field regardless of whether we correct the mean distance.

\subsection{Strategy 3: Placing Galaxies at Redshift Coordinates (Recommended)}
\label{sec:velocity_redshift}

The most robust approach is to place galaxies at their \emph{redshift-space} coordinates $s = cz/H_0$, rather than attempting to correct inferred distances. This yields the estimator
\begin{equation}
\vobs(s) = s - d\, ,
\end{equation}
where each galaxy is assigned its velocity estimate $s - d$ but positioned at $s$.

To assess the bias in this estimator, we compute $\conm{d}{s}$ using the law of total expectation:
\begin{equation}
\conm{d}{s} = \int \conm{d}{r}\,P(r|s)\,{\rm d}r\, .
\end{equation}
For an unbiased distance estimator (iTF with lognormal correction, or fTF after $\SFTF$ correction), $\conm{d}{r} = r$, giving
\begin{equation}
\conm{d}{s} = \int r\,P(r|s)\,{\rm d}r = \conm{r}{s}\, .
\end{equation}

The conditional distribution is
\begin{equation}
P(r|s) \propto P(s|r)\,P(r) \propto \exp\!\left[-\frac{(s-r-V(r))^{2}}{2\sigma_v^{2}}\right] r^{2}n(r)S(r)\, ,
\end{equation}
where $\sigma_v$ represents the small-scale velocity dispersion and $S(r)$ accounts for observational selection. Expanding about the real-space solution $r_t$ defined by $s = r_t + V(r_t)$, and assuming $|V'(r_t)| \ll 1$, we obtain
\begin{equation}
\conm{r}{s} \approx r_t + \frac{\sigma_v^{2}}{r_t}\,\frac{{\rm d}\ln P(r)}{{\rm d}\ln r}\bigg|_{r_t}\, .
\end{equation}

Therefore, the mean observed velocity at redshift coordinate $s$ is
\begin{equation}
\conm{\vobs}{s} = s - \conm{d}{s} \approx V(r_t) - \frac{\sigma_v^{2}}{r_t}\,\frac{{\rm d}\ln P(r)}{{\rm d}\ln r}\bigg|_{r_t}\, .
\end{equation}

The bias term $\propto \sigma_v^{2}/r_t$ is small for typical surveys where $r_t \gg \sigma_v/H_0 \sim 20\,h^{-1}{\rm Mpc}$. Moreover, this bias is uncorrelated with the large-scale velocity field itself, since it depends only on the density gradient and velocity dispersion. For surveys extending to $\sim100\,h^{-1}{\rm Mpc}$, this approximation introduces errors below the $\sim10\%$ level on large scales, making it the preferred approach for velocity field reconstruction. The redshift-space assignment minimizes bias on scales large compared to $\sigma_v/H_0$, avoids vMB from distance scatter, does not require detailed knowledge of the underlying density field $n(r)$, and provides a practical and robust reconstruction suitable for large-scale structure analyses.

\subsection{Strategy 4: Using PDFs and Bayesian Methods}
\label{sec:VfromPDF}

A more sophisticated approach exploits the joint probability of distance indicator $d$, true distance $r$, and observed redshift coordinate $s$:
\begin{equation}
P(r,d,s) = P(r)\,P(d|r)\,P(s|r,V)\, ,
\end{equation}
where the likelihood factors are modeled as
\begin{equation}
P(s|r,V) \propto \exp\!\left[-\frac{(s-r-V)^{2}}{2\sigma_{v}^{2}}\right].
\end{equation}

Maximizing with respect to $V$ yields $V_{\max} = s - r$, as expected. Maximizing with respect to $r$ gives
\begin{equation}
\frac{{\rm d}\ln P(r)}{{\rm d}r} - \frac{\ln(r/d)}{\sigma_{d}^{2}r} + \frac{s-r-V}{\sigma_{v}^{2}} = 0\, .
\end{equation}
Substituting $V = s - r$ eliminates the last term, yielding
\begin{equation}
\frac{{\rm d}\ln P(r)}{{\rm d}\ln r} - \frac{1}{\sigma_{d}^{2}}\ln(r/d) = 0\, ,
\end{equation}
which determines $r_{\max}$ from $P(d|r)P(r)$, independent of $s$. The velocity estimate is then $\ves = s - r_{\max}$.

If instead a velocity model $V(r)$ is assumed, the likelihood becomes
\begin{equation}
P(s|r) \propto \exp\!\left[-\frac{(s-r-V(r))^{2}}{2\sigma_{v}^{2}}\right],
\end{equation}
and maximization yields
\begin{equation}
\frac{{\rm d}\ln P(r)}{{\rm d}\ln r} - \frac{1}{\sigma_{d}^{2}}\ln\frac{r}{d} + \frac{r}{\sigma_{v}^{2}}\,[s-r-V(r)]\,[1+V'(r)] = 0\, .
\label{eq:MAP_r_dep_V}
\end{equation}
In the limit $\sigma_v/\sigma_d \ll 1$ with finite $\sigma_v$, the solution is heavily weighted toward the redshift constraint $s = r + V(r)$, recovering the result from \S\ref{sec:velocity_redshift}. This convergence demonstrates that Bayesian methods, when properly accounting for the relative uncertainties in distance and redshift measurements, naturally favor redshift-based coordinate assignments for velocity field reconstruction.

In Appendix~\ref{sec:graziani} we discuss an application of the Gibbs sampling algorithm by \citet{Graziani_Hoffman} and show that it suffers from biases.

\section{Illustration Using a Toy Model}
\label{sec:toy_model}

To illustrate the reconstruction of velocity fields from distance and velocity estimators, we employ a spherically symmetric toy model that isolates systematic effects while avoiding stochastic fluctuations inherent to $N$-body simulations with finite tracer populations. 

The model features a single overdensity centered at distance $r_0 = 80$ Mpc from the observer. For a galaxy at true distance $r$ from the observer along a line of sight at angle $\theta$ relative to the cluster radius vector (where $\theta = 0$ points directly through the cluster center), the distance from the cluster center is $r_{\rm rel}(r,\theta) = \sqrt{r_0^2 + r^2 - 2r_0 r\cos\theta}$.

The number density field depends only on the distance from the cluster center: $n(r_{\rm rel}) = 1 + \delta(r_{\rm rel})$, where the density contrast relative to the mean is
\begin{equation}
\delta(r_{\rm rel}) = \frac{1}{1 + (r_{\rm rel}/r_s)^{\beta(r_{\rm rel})}}\, , \quad
\beta(r_{\rm rel}) = 2 + \frac{r_{\rm rel}}{r_s}\, ,
\label{eq:delta_toy}
\end{equation}
with $r_s = 1$ Mpc. This analytic profile produces a centrally concentrated overdensity that steepens smoothly with radius, emulating realistic large-scale structures.

Galaxies are assigned absolute magnitudes according to the 2MASS $K$-band luminosity function   \citep{BND12}. The peculiar velocity field is computed from linear gravitational instability theory. The radial velocity at distance $r_{\rm rel}$ from the cluster center (positive outward) is
\begin{equation}
V(r_{\rm rel}) = -\frac{f}{r_{\rm rel}^2}\int_0^{r_{\rm rel}} \delta(u)\,u^2\,{\rm d}u\, ,
\label{eq:Vrrel}
\end{equation}
where $f = 1$ is the linear growth rate. The line-of-sight velocity component at observed distance $r$ is the projection of the velocity at the corresponding $r_{\rm rel}(r,\theta)$ onto the line of sight: $V_{\rm los}(r,\theta) = V(r_{\rm rel})\,\cos\psi$, where $\cos\psi$ is the cosine of the angle between the outward radius vector from the cluster center and the line-of-sight direction.

To simulate distance-indicator measurements, we add lognormal scatter to the true distances. For each galaxy at true distance $r_{\rm true}$ from the observer, the observed distance $d_{\rm est}$ is drawn from $\ln d_{\rm est} = \ln r_{\rm true} + \epsilon$, where $\epsilon \sim \mathcal{N}(0, \sigma_{\ln}^2)$ with $\sigma_{\ln} = (\ln 10/5)\sigma_\mu$ and $\sigma_\mu = 0.3$ mag (typical of the Tully-Fisher relation). The redshift-space coordinate is $s = r_{\rm true} + V_{\rm los}(r_{\rm true},\theta)/H_0$.

This framework allows us to generate mock catalogs with known true distances, observed distances, and redshift-space coordinates, enabling controlled tests of velocity field reconstruction methods in the presence of realistic distance errors and spatial selection effects arising from magnitude-limited observations.

\subsection{Results from the toy model}
We place the center of the perturbation \cref{eq:delta_toy} at $r_0=80$ Mpc and draw an ensemble of values for true distances, $r$, sampled from the 
real space density distribution, $r^2[1+\delta(r_\textrm{rel})]$, in a line of sight passing through the perturbation center, out to a maximum distance of $r=200$ Mpc. 
We then assign each point in the ensemble an absolute magnitude, $M$, according to the 2MRS luminosity function and 
compute the corresponding apparent magnitudes using the true 
distances. The sample is then trimmed to create a magnitude-
limited catalog by imposing the limit $m_l=12$. The absolute 
magnitudes are used to derive linewidth parameters $\eta$ in the iTF relation with $\gamma=0.12$ and $\sigma_\eta=0.07$. 
The number of points in the magnitude-limited sample is 
$\approx 2\times 10^4$, which is unrealistically large; 
however, we aim at demonstrating systematic dMB and vMB 
inherent to magnitude-limited surveys, not at assessing random 
errors. The large number is thus appropriate as it emphasizes 
the systematic biases while suppressing statistical 
fluctuations.

Given true distances, $r$, the iTF is used to derive observed distances, $d$. In all results presented here, the correction \cref{eq:EITF} due to the lognormal nature of the iTF scatter has been incorporated.  

Furthermore, for each $d$ we draw random samples, $\hat r$, from the distribution $P(r|d)$ (cf. \cref{eq:Prdconditf}). We then compute the mean $\bar r(d)$ using these samples. Alternatively, we can compute $\bar r(d)$ using the FLS method described in \S\ref{sec:LS}. Both yield similar results and we use the mean of $P(r|d)$. Since $n(r)$ is unknown, a direct estimation of $P(r|d)$ from observations is not possible and only $\bar r(d)$ can in principle be estimated from observations using the FLS method. Nonetheless, our goal here is to illustrate that vMB is inherent in velocity field extraction even if we use a full sampling of $P(r|d)$.

\begin{figure}[ht!]
\plotone{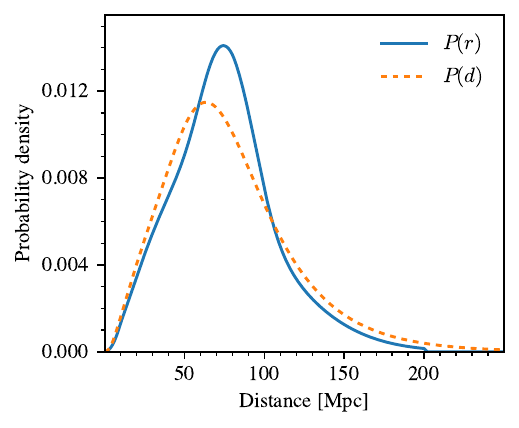}
\caption{Normalized density distributions along a line of sight through the overdensity center at $r_0 = 80$ Mpc. 
The solid line shows the true distance distribution $p(r)$, while the dashed line shows the observed (iTF) distance distribution $p(d)$ from the toy model catalog. 
The decline at large distances reflects the imposed magnitude limit.
\label{fig:N_los}}
\end{figure}

The distribution of true, $r$, and observed, $d$, distances is plotted in \cref{fig:N_los}. The magnitude limit reduction in the number of points is evident at larger distances. The distribution $P(d)$ is not only smeared compared to $P(r)$ but it is also skewed to the left relative to the peak in $P(r)$. Further, $P(d)$ extends beyond the maximum $r=200$ Mpc.  

\begin{figure}[ht!]
\epsscale{1.2}
\plotone{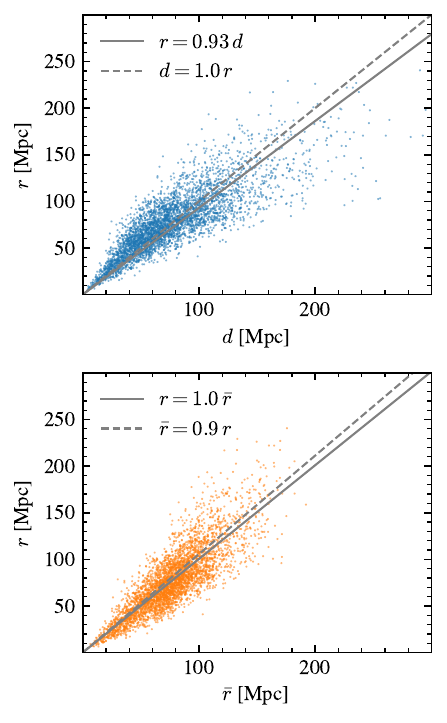}
\caption{Upper panel: true distances $r$ versus observed iTF distances $d$. 
Lower panel: true distances $r$ versus $\bar r$. 
Solid lines: regression of $r$ on the distance estimator. 
Dashed lines: regression of the estimator on $r$.
\label{fig:dist_two_pane}}
\end{figure}

The top panel 
in \cref{fig:dist_two_pane} shows a point-by-point comparison 
between $r$ and $d$ for a random subsample of points. The 
constant iTF scatter in $\ln(r/d)$ translates into a spread in 
$r-d$ that increases linearly with distance as seen in the 
figure. It is immediately evident from the points with $d>200$ 
Mpc that the mean of $r$ at a given $d$ is not equal to $d$. This is confirmed by the slope of the linear regression line (solid) of $r$ on $d$ which yields a slope below unity. The 
slope of the regression of $d$ on $r$ (dashed) is however 
unity, indicating that the iTF $d$ is unbiased in the sense 
$\conm{d}{r}=r$ (modulo the correction in \cref{eq:EITF}).

The bottom panel of \cref{fig:dist_two_pane} shows a scatter 
plot of $r$ vs $\bar r(d)$ for all points in the same subsample 
as in the top panel. If $\bar r(d)$ is a one-to-one mapping, as 
is the case in the toy model, then points with similar $\bar r$ 
in the figure correspond to similar values of $d$. Therefore, 
values of $r$ for any small range of $\bar r$ should scatter 
randomly around this $\bar r$. This is indeed the case and is 
demonstrated by the fact that the regression of $r$ on $\bar r$ 
is unity as indicated in the figure by the solid line, in 
contrast to the regression of $\bar r$ on $r$ which deviates 
from unity.

%%%%%%%%%%%%%%%%%%%%%%  VELOCITY PLOTS NOW

\begin{figure*}[ht!]
\epsscale{1.2}
\plotone{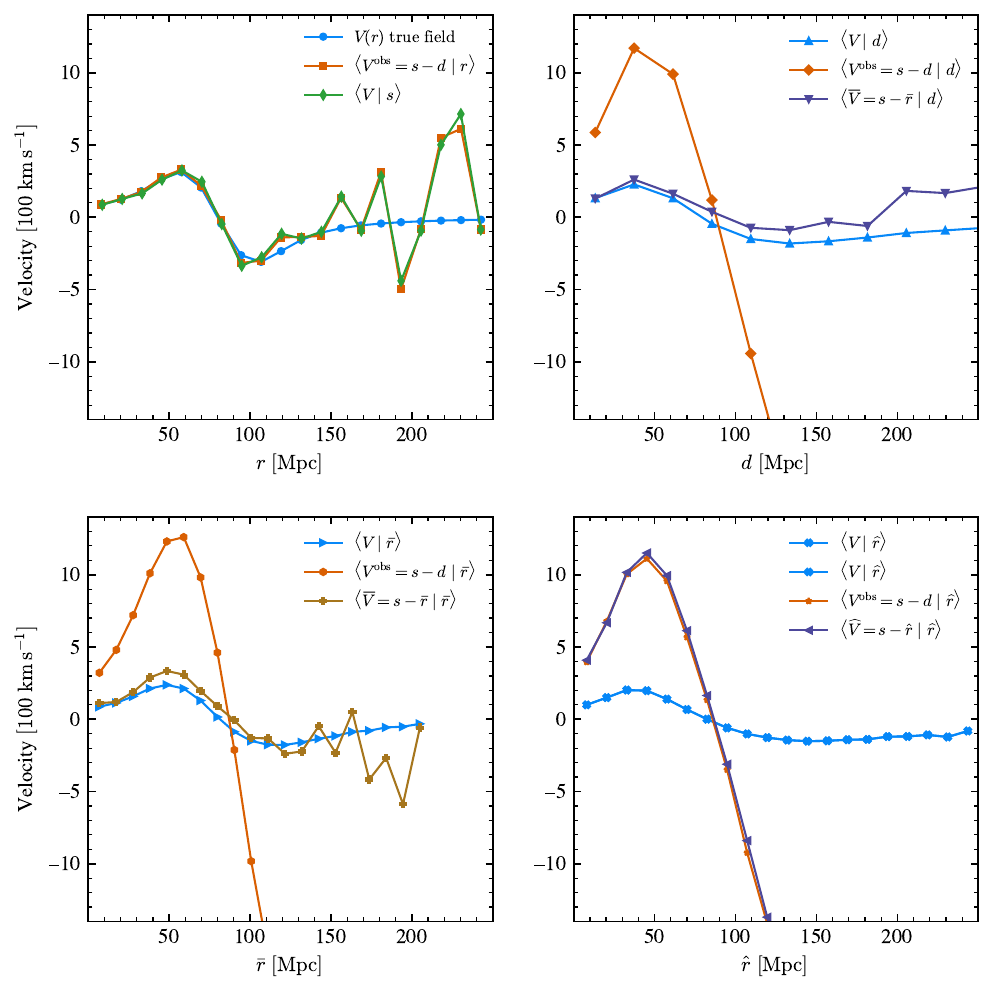}
\caption{Various line-of-sight velocity estimators as functions of different distance estimators for the toy model magnitude-limited catalog. 
Each panel shows binned mean velocities  in distance binns.
Top-left: velocities vs true distance $r$, including $V^{\rm obs}$ vs redshift-space distance $s$ (green).
Top-right: velocities vs observed iTF distance $d$.
Bottom-left: velocities vs posterior mean distance $\bar r$, with the corresponding velocity estimator $\bar V = s - \bar r$.
Bottom-right: velocities vs sampled distance $\hat r$ from $P(r|d)$, with $\hat V = s - \hat r$.
\label{fig:V_binned_panels}}
\end{figure*}

Velocity field estimators along a single line of sight are plotted in \cref{fig:V_binned_panels}. 
Each panel shows velocities binned by a different distance variable, allowing direct comparison 
of biases inherent to each choice.

\textit{Top-left: velocities binned in true distance $r$ and redshift $s$.}
The blue curve shows the actual line-of-sight velocity as a function of true distance, $V(r)$, 
derived from the linear theory relation \cref{eq:Vrrel}; this is the ground truth we wish to 
recover from our ``observations.'' The orange curve represents the observed velocity versus 
true distance, $\vobs(r)=\conm{\vobs=s-d}{r}$ (cf.~\S\ref{sec:velocity_at_d}), which is unbiased 
and simply a noisy version of $V(r)$. The green curve shows $\vobs(s)$ 
(cf.~\S\ref{sec:velocity_redshift}) binned in redshift $s$; to first order $\vobs(s)\approx \vobs(r)$, 
explaining the proximity of these two curves.

\textit{Top-right: velocities binned in observed distance $d$.}
The blue curve, $\conm{V}{d}$, obtained by averaging over points in bins of observed distance $d$. Since $\conm{V}{d}\ne V(r=d)$ the blue curve does not coincide with, and is actually a biased version of,  the blue curve in the top-left panel showing  true field. The orange curve, $\vobs(d)$, is 
strongly biased: since points with $d>r$ have $\vobs=s-d=V(r)+r-d<V(r)$ and vice versa, this field 
exhibits spurious velocity patterns even when $V=0$ everywhere. The dark blue curve represents 
$\bar{V}=s-\bar{r}$ in bins of $d$; since $\hat{r}$ samples $r$ given $d$, this curve coincides 
with $\hat{V}=s-\hat{r}$ versus $d$ and is likewise a biased representation of $V(r)$.

\textit{Bottom-left: velocities binned in $\bar{r}$.}
The conditional true velocity $\conm{V}{\bar r}$ (blue curve) is mildly biased with respect to $V(r)$ (blue curve in top-left panel), as explained 
in \S\ref{sec:velocity_at_rbar}. The function $\bar{V}(\bar{r})$ lies close to $\conm{V}{\bar{r}}$ because 
$\bar{r}(d)$ in the toy model is a one-to-one function, so the mean of $r$ values in bins of $\bar{r}$ 
equals $\bar{r}(d)$ in accordance with \cref{eq:second_impl}. The field $\vobs(\bar{r})$ is strongly biased, effectively equivalent to 
$\vobs(d)$ due to the one-to-one mapping between $d$ and $\bar{r}(d)$.

\textit{Bottom-right: velocities binned in $\hat{r}$.}
The blue curve $\conm{V}{\hat{r}}$ is mildly biased and of lower amplitude than the ground truth $V(r)$ plotted in the top-left panel. 
This bias arises because $\hat{r}$ is a fair sampling of $r$ given $d$, not given the actual true $r$; 
Appendix~\ref{sec:rhatr} provides a calculation of $P(\hat{r}|r)$. The field 
$\hat{V}(\hat{r})=\conm{s-\rh}{\rh}=\conm{\rt-\vt}{\rh}-\rh$ (dark blue) is biased because, as 
demonstrated in the appendix, $\conm{\rt}{\rh}\ne\rh$. The orange curve $\vobs(\rh)$ is also biased 
since $\conm{d}{\rh}=\rh$ (because $P(d|\rh)$ equals the PDF of $d$ given $r_{\rm true}$), which 
explains why this curve closely follows the dark blue $\hat{V}(\hat{r})$ curve.

%%%%%%%%%%%%%%%%%%%%% 
 \section{Machine Learning and Modified Wiener Filtering for vMB Mitigation}
\label{sec:ml_wiener}
Machine learning offers a potential route to mitigate vMB. An autoencoder trained on mock catalogs to map the observed velocity field $\vobs(d)$ in distance space to the true field $\vtru(r)$ in real-space coordinates would, in principle, learn to estimate the conditional expectation $\langle \vtru(r) | \vobs(d) \rangle$. This estimator is unbiased in the sense that averaging over all realizations of the true field that are consistent with the observations recovers the correct mean. In the limit of Gaussian fields, a well-trained autoencoder converges to the optimal linear Wiener Filter (WF) \citep{Punya2023,Lilow2024}, so examining the WF provides insight into the fundamental limitations and potential of ML-based approaches.

The Wiener filter has been employed to recover velocity fields in several studies \citep[e.g.,][]{Hoffman15}. However, these standard implementations fundamentally fail to address vMB because they make a critical simplifying assumption: they place galaxies at their observed distances $d$ (or at a weighted average between redshift and distance  \citep{Hoffman2021}) and then treat each velocity measurement as a noisy version of the true velocity \emph{at that prescribed observed distance}. 

This approach ignores the essential physics of vMB: a galaxy observed at distance $d$ does not sample the velocity at a single location $d$, but rather samples velocities from a \emph{distribution} of true distances $r$ weighted by $P(r|d)$. The standard WF thus implicitly assumes
\begin{equation}
\vobs(d) \approx \vtru(d) + \text{noise}\, ,
\end{equation}
which neglects the systematic bias inherent in \cref{eq:vobsd_bias}. The correlations used in standard WF are simply the ordinary velocity-velocity correlations $C_V(d_i, d_j)$ evaluated at the observed distances, with no accounting for the fact that $\vobs(d)$ is fundamentally related to $\vtru(r)$ through an integration along the line of sight.

We present here a modification of the WF that explicitly addresses vMB by properly marginalizing over the distance uncertainty encoded in $P(r|d)$. The key conceptual difference is that we seek the conditional mean field $\conm{\vtru(\vr)}{\vobs(d)}$, recognizing that $\vobs(d)$ is related to the underlying field through \cref{eq:vobsd_bias}.

The general WF estimate for the desired velocity field \citep{hr91, 1999ApJ...520..413Z} has the form
\begin{equation}
\vwf(\vr) = \sum_{i,j} \zeta_i(\vr)\, \Xi^{-1}_{ij}\, \vobs(\pmb{d}_j)\, ,
\end{equation}
where $\pmb{d}_i$ denotes the observed coordinates (angular position and distance $d_i$) of galaxy $i$, and the sum extends over all galaxies in the survey. The cross-correlation $\zeta$ and auto-correlation $\Xi$ are now defined to properly account for vMB:
\begin{equation}
\zeta_i(\vr) = \conm{\vobs(\pmb{d}_i)\,\vtru(\vr)}{}
\end{equation}
and
\begin{equation}
\Xi_{ij} = \langle{\vobs(\pmb{d}_i)\,\vobs(\pmb{d}_j)}\rangle\, .
\end{equation}

These correlations differ \emph{critically} from the standard WF. Rather than evaluating velocity correlations at the observed positions, $\zeta_i(\vr)$ and $\Xi_{ij}$ properly account for the fact that a galaxy at observed distance $d$ samples velocities from a distribution of true distances weighted by $P(r|d)$, thereby marginalizing over the distance uncertainty that causes vMB.

To compute the correlations, we use \cref{eq:vobsd_bias} with $P(r|d)$ from \cref{eq:Prdcondftf}. For a galaxy at observed distance $d_i$ along direction $\nh_i$, the cross-correlation becomes
\begin{equation}
\zeta_i(\vr) = \int w(r|d_i)\, C_V(r\nh_i, \vr)\, \dd r\, ,
\label{eq:zeta_vMB}
\end{equation}
where $C_V(\vr_1, \vr_2) = \langle{\vtru(\vr_1)\vtru(\vr_2)}\rangle$ is the velocity correlation function, and the weight function is
\begin{equation}
w(r|d) = \frac{r^2 n(r) \glnf{r}{d}}{\int r'^2 n(r') \glnf{r'}{d}\, \dd r'}\, ,
\end{equation}
with $n(r)$ the spatial number density and $\gln$ the log-Gaussian kernel describing distance scatter (Eq.~\ref{eq:Prdcondftf}). Similarly, the auto-correlation becomes
\begin{equation}
\Xi_{ij} = \iint w(r|d_i) w(r'|d_j)\, C_V(r\nh_i, r'\nh_j)\, \dd r\dd r'\, .
\end{equation}

The line-of-sight integrals in \cref{eq:zeta_vMB} encode the essential ingredients  missing from standard implementations: they properly average the velocity correlation function over the probability distribution $P(r|d)$, thus accounting for the fact that observed velocities at distance $d$ reflect contributions from a range of true distances. Similar expressions can be obtained for the iTF.

%\subsection{Implementation and Limitations}

We note that WF uses only the statistical properties of the underlying fields.  
Therefore, although the modified WF removes the systematic bias caused by the mismatch between $r$ and $d$, it does not guarantee an unbiased reconstruction in any specific region; WF returns the minimum--variance linear estimator, not necessarily a point-by-point unbiased map.

\subsection{Illustration of Modified WF relative to other fields using a particle-mesh simulation}
\label{sec:PM}

As a preliminary illustration, we consider a simplified setup with Gaussian, fixed-amplitude distance errors and the distant–observer limit. This is a simplified illustration and is not an extension of the toy-model test  in \S\ref{sec:toy_model}.

To explore how different field estimators behave in a 3D cosmological setting, we performed a particle-mesh simulation of the Planck cosmology in a $512^3$ cubic grid of length $1000\hmpc$ on a side, with $512^3$ particles. We assumed a Gaussian distance error of $\sigma_d = 10\hmpc$ and adopted the distant observer limit with the line of sight in the $z$-direction. We construct four velocity field estimates: the true field $V(r)$ in real-space coordinates; the observed field $\vobs(d)$ placed at observed (biased) distances $d$; the field $\vobs(s)$ placed at redshift coordinates $s = r + v_r(r)$, representing our advocated coordinate choice; and the modified Wiener-filtered field $\vwf(r)$ from \S\ref{sec:ml_wiener}. 

\begin{figure*}[ht!]
\epsscale{1.}
\plotone{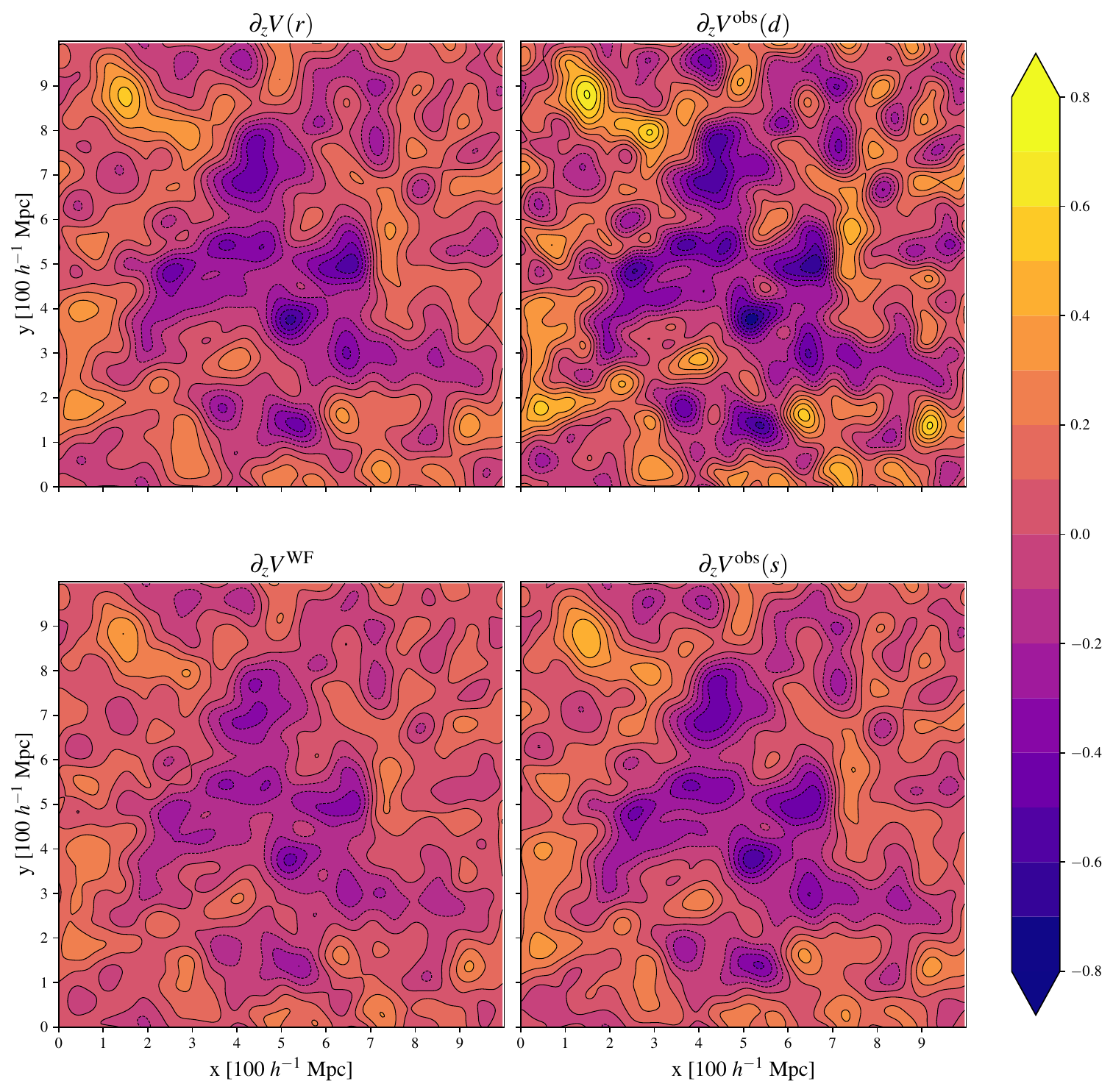}
\caption{
 Spatial derivatives of the velocity field, 
$\partial V_z / \partial z$ in the true, observed, and redshift--space 
representations (top and bottom--left)
All fields were smoothed with a Gaussian kernel of width 
$20~h^{-1}\,\mathrm{Mpc}$. 
Mock distance errors with a Gaussian scatter of 
$10~h^{-1}\,\mathrm{Mpc}$ were applied to generate the observed and 
redshift--space quantities.
\label{fig:pm_sim}}
\end{figure*}

\Cref{fig:pm_sim} shows heat maps of the $z$-derivative of the line-of-sight velocity field in the same $x$--$y$ slice for all four fields. The large differences between the true $V(r)$ (top-left) and $\vobs(d)$ (top-right) that we saw in the toy model (\S\ref{sec:toy_model}) appear here as well. The fluctuations in $\vobs(d)$ are evidently enhanced compared to the true field, arising because galaxies scattered to larger (smaller) distances by measurement errors tend to come from denser (less dense) regions, artificially steepening velocity gradients. The modified WF field (bottom-left) exhibits substantial suppression of fluctuations. Most remarkably, $\vobs(s)$ (bottom-right), obtained by placing observed velocities at redshift coordinates, closely resembles the true field in both spatial structure and amplitude, validating our central thesis that coordinate choice is crucial for unbiased velocity field reconstruction.

\begin{figure}[ht!]
\plotone{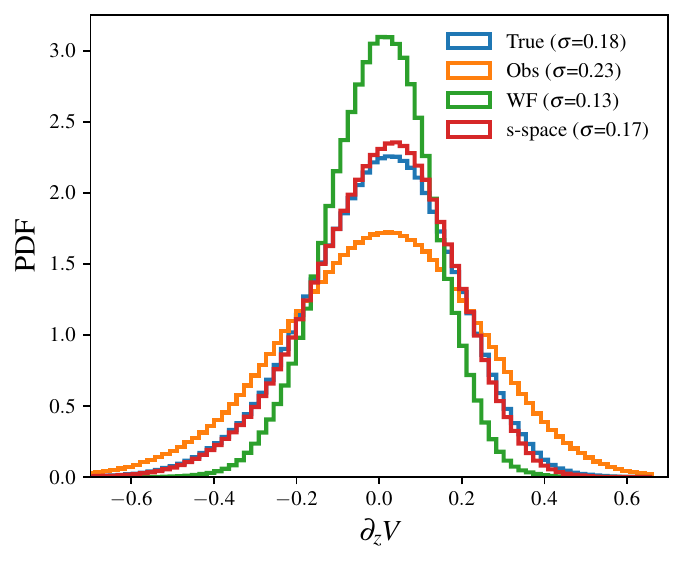}
\caption{
Probability density functions of $\partial_z V$ for the four 
three--dimensional fields shown in the previous figure. 
Values of the standard deviations, $\sigma$, are listed in the figure. 
The comparison shows the enhancement of the fluctuations in 
$V^{\mathrm{obs}}(d)$ (orange) and the suppression in the
Wiener--filtered field $V^{\mathrm{WF}}$ (green), while the redshift--space field 
$V^{\mathrm{obs}}(s)$ (red) is close to the true field $V(r)$ (blue). 
\label{fig:pm_sim_pdf}}
\end{figure}

\Cref{fig:pm_sim_pdf} shows the PDFs of $\partial v_r/\partial z$ corresponding to the fields, quantifying these visual impressions. The true field (blue) and $\vobs(s)$ (red) have nearly identical widths with $\sigma = 0.18$ and $0.17$ respectively, demonstrating that placing galaxies at redshift coordinates preserves the statistical properties of the velocity field. In contrast, $\vobs(d)$ (orange) shows $\sigma = 0.23$, a $\sim 30\%$ enhancement quantifying the systematic bias from vMB when using distance-corrected positions. The WF (green) yields $\sigma = 0.13$, substantially suppressed as it trades amplitude for noise reduction.
%%%%%             modified WF
% generated by /Users/adi/Cosmological-Particle-Mesh-Simulation-master/src/analyze_for_Malmquist.ipynb

%%%%%%%%%% HISTOGERAM

\section{Summary and discussion}
\label{sec:discussion}
In the standard paradigm for structure formation, the equivalence principle guarantees that galaxies and dark matter share the large-scale peculiar velocity field. Consequently, the galaxy velocity field is directly related to the underlying mass density through gravitational instability theory.

Observationally, peculiar velocities are inferred by combining redshift measurements with independent distance estimates. For spiral galaxies, the TF relation between luminosity and rotational velocity  serves as the primary distance indicator. Given a measured redshift $cz$ and an inferred distance $d$, the line-of-sight peculiar velocity is simply $v = cz - H_0 d$. However, this apparently straightforward procedure is fraught with systematic biases that can severely compromise large-scale velocity field reconstructions.

Because of the intrinsic scatter in these correlations, the linear forward TF relation $M(\eta)$, which predicts absolute magnitude from rotational linewidth, and the inverse relation $\eta(M)$, which predicts linewidth from magnitude, are not exact inverses of one another. The inverse TF yields distance moduli that are unbiased for galaxies with the same true distance, while the forward relation introduces a systematic bias represented by a multiplicative correction factor $\mathcal{S}^{\mathrm{frw}}$ that depends on the survey selection function. We demonstrate that this factor can be estimated directly from the observed data whenever the catalog is magnitude-limited and the selection on $\eta$ is weak. After this correction, both formulations yield consistent and unbiased peculiar velocity estimates \emph{at fixed true distance}. However, true distances are unknown for individual galaxies, and therefore an estimator that is unbiased at fixed $r$ cannot be used to assign unbiased velocities to individual objects. 

This brings us to a point that has not been sharply articulated in the literature: the difference between \emph{distance} Malmquist bias (dMB), which affects individual distance estimates, and \emph{velocity} Malmquist bias (vMB), which corrupts reconstructed velocity fields. While dMB can in principle be approximately corrected using methods such as the Feast--Landy--Szalay prescription (cf.\ SW95), vMB persists even after such corrections. The reason is fundamental: vMB originates from the \emph{scatter} in the distance estimates rather than from any systematic offset in their mean. A galaxy observed at distance $d$ may have come from any true distance $r$ within the posterior $P(r\mid d)$, and this spread in $r$ propagates directly into the velocity field regardless of whether the mean of $P(r\mid d)$ equals $d$ or not.
We have demonstrated this through both analytical arguments and a spherically symmetric toy model. When galaxies are placed at their inferred distances (whether raw, bias-corrected, or sampled from the posterior) the resulting velocity field exhibits systematic distortions. 

The solution we advocate is conceptually simple: place galaxies at their observed redshift coordinates $s = cz/H_0$ rather than at inferred distances, as has been done by \citep{Aaronson1982}. Since redshifts are unaffected by distance indicator errors, this assignment avoids the mixing of velocities from different true positions that causes vMB. The cost is a residual bias 
over scales of order  $\sigma_v/H_0$, where $\sigma_v$ is the small-scale incoherent velocity dispersion. Only for a relatively small number of galaxies  the distance error is smaller than this scale. For those one can indeed place galaxies at the observed distance directly inferred from the TF relation.  As an example, for $\sigma_v\approx 250\kms$ and a $10\% $
distance indicator error, it is preferred to use observed distances only for the subset of galaxies nearer than $15\hmpc$.

This result has direct implications for velocity--gravity comparisons, which constrain the growth rate $f\sigma_8$ by comparing observed velocities with predictions from the gravitational field inferred from redshift surveys. Our analysis indicates that predicted velocities should be evaluated at the redshift coordinates of peculiar velocity tracers, not at their inferred distances. As shown in Appendix~\ref{sec:vgcomp}, this approach yields growth rate constraints that are unbiased to leading order. 

We have also explored  a more sophisticated approache, including a modified Wiener filter that properly marginalizes over the distance uncertainty encoded in $P(r|d)$. This formalism  yields correlations that account for the fact that observed velocities at distance $d$ sample a distribution of true distances. While the modified filter removes systematic bias, it necessarily suppresses amplitude to reduce noise- the price paid by any linear estimator operating on scattered data. Whether machine learning methods can improve on this remains an open question; we emphasize  that in the Gaussian limit, well-trained autoencoders converge to the Wiener filter, suggesting that the fundamental limitations may be difficult to circumvent.

%%%%%%%%%%%%%%%%%%%%  ORIGINAL

\section{Acknowledgments}

 This research has been supported by a grant (\#893/22) from the Israel Science Foundation and a grant from the Asher Space Research Institute.

%\software{astropy %\citep{2013A&A...558A..33A,2018AJ....156..123A,2022ApJ...935..167A},  
%          Cloudy \citep{2013RMxAA..49..137F}, 
%          Source Extractor \citep{1996A&AS..117..393B}
%          }


\appendix

\newcommand{\efunc}[1]{\mathrm{e}^{#1}}
\newcommand{\gfunc}[2]{\mathrm{e}^{-\frac{#1^2}{2#2^2}}}
\def\dd{\mathrm{d}}
\def\e{\mathrm{e}}
\def\mue{\mu_\textrm{e}}
\def\mud{\mu_\textrm{d}}
\def\re{r_\textrm{e}}
\def\mur{\mu_r}
\def\rhoFP{\widehat \rho}
\def\ihat{\widehat i}

\section{Fundamental Plane as a distance indicator}
\label{sec:fundamental}

The Fundamental Plane (FP) relates an elliptical galaxy's effective radius $R_e$, central velocity dispersion $\sigma_0$, and surface brightness $I_e$:
\begin{equation}
    \rho \equiv \log R_e = a\, i + b\, s + c \;, \quad i \equiv \log I_e, \quad s \equiv \log \sigma_0 \;.
\end{equation}
Since $\sigma_0$ and $I_e$ are distance-independent while $R_e = \theta_e d_A$, comparing the predicted $R_e$ to the observed angular radius $\theta_e$ yields the distance.

With $t = \log\theta_e$ and apparent magnitude $m = -2.5\log f$, the surface brightness relation $I \propto f/\theta^2$ gives
\begin{equation}
    t = -\frac{m}{5} - \frac{i}{2} \;, \qquad
    \rho = t + \mu_r = -\frac{m}{5} - \frac{i}{2} + \mu_r \;,
\end{equation}
where $\mu_r = 5\log r + \mathrm{const}$ is the true distance modulus.

\subsection{Forward formulation}

The forward method predicts $\rho$ from the distance-independent observables $(i,s)$. Writing $P(\rho,i,s) = P(\rho|i,s)\,P(i|s)\,P(s)$ with
\begin{equation}
    P(\rho|i,s) \propto \gfunc{\rho - \rhoFP(i,s)}{\Delta_\rho} \;, \qquad \rhoFP(i,s) = a i + b s \;,
\end{equation}
the estimated distance modulus is $\mu_d = \rhoFP(i,s) + m/5 + i/2$.

Including the selection function $S(m,i,s)$ and spatial density $n(r)$, the joint probability $P(r,\rho,i,s,m) \propto r^2 n(r)\,P(\rho,i,s)\,S(m,i,s)$ integrates to
\begin{equation}
    P(r|d) = \frac{r^2 n(r)\, e^{-(\mu_r-\mu_d)^2/2\Delta_\rho^2}}
    {\int \dd r\; r^2 n(r)\, e^{-(\mu_r-\mu_d)^2/2\Delta_\rho^2}} \;.
\end{equation}

\subsection{Inverse formulations}

Writing $P(\rho,i,s) = P(i|\rho,s)\,P(\rho|s)\,P(s)$ with $P(i|\rho,s) \propto e^{-(i-\hat{i})^2/2\Delta^2}$ and $\hat{i} = a\rho + bs$, one obtains after marginalizing over the selection function:
\begin{equation}
    P(r|\mu_d) = \frac{r^2 n(r)\,\mathcal{S}(r)\, e^{-(\mu_r-\mu_d)^2/2\Delta^2}}
    {\int \dd r\; r^2 n(r)\,\mathcal{S}(r)\, e^{-(\mu_r-\mu_d)^2/2\Delta^2}} \;,
\end{equation}
where $\mathcal{S}(r) = \int \dd m\,\dd s\; P(\mu_r - 0.2m - i/2 | s)\,S(m,s)$.

\subsubsection{Predicting $s$ from $(i,\rho)$}

Alternatively, $P(i,s,\rho) = P(s|i,\rho)\,P(i,\rho)$ with $P(s|i,\rho) \propto e^{-(s-\hat{s})^2/2\Delta_s^2}$ and $\hat{s} = (\rho - ai)/b$ yields
\begin{equation}
    P(r|\mu_d) = \frac{r^2 n(r)\,\widetilde{\mathcal{S}}(r)\, e^{-(\mu_r-\mu_d)^2/2(b\Delta_s)^2}}
    {\int \dd r\; r^2 n(r)\,\widetilde{\mathcal{S}}(r)\, e^{-(\mu_r-\mu_d)^2/2(b\Delta_s)^2}} \;,
\end{equation}
with $\widetilde{\mathcal{S}}(r) = \int \dd m\,\dd i\; P(i, \mu_r - 0.2m - i/2)\,S(m,i)$.

In all cases, the conditional $P(r|d)$ takes the standard form involving a Gaussian in $\mu_r - \mu_d$ weighted by $r^2 n(r)$ and a selection-dependent factor, paralleling the Tully--Fisher analysis.

%%%%%%%%%%%%%%%%%%%%%%%%%%%%%%%%%%%    SFTF
\section{Direct determination of $\SFTF$ from the observed distribution of objects}
\label{sec:SFTF}
For a catalog with a strict magnitude limit $m_\text{l}$, it is possible to compute $\SFTF$ directly from the data by adapting the method of \citet{dav82}, originally developed to estimate the selection function of magnitude-limited redshift surveys of galaxies.

The method is applicable for a magnitude limited catalog and with a possible selection criteria imposed on $\eta$ independent of the magnitude, i.e. $S(m,\eta) = S_\eta(\eta)$ for $m<m_l$ and zero otherwise. 
Therefore \cref{eq:SFTF} becomes
\begin{equation}
\SFTF = \int_{\eta_\textrm{l}} \dd \eta  \, \phi(\eta)S_\eta(\eta) \; ,
\end{equation}
where $\eta_{l}(d) = (m_{l} - \muFTF - b)/a$ ($a<0$).
This expression is analogous to the fraction of galaxies brighter than a magnitude limit, except that $\phi(\eta)$ replaces the luminosity function.

The number of objects actually observed with $\dFTF < d$ but that would remain observable at distances $\ge d$ is
\[
T(d) = \frac{4\pi}{3} d^{3} \SFTF(d) \; .
\]
These correspond to objects with $\eta \ge \eta_{l}(d)$.
Similarly,
\[
F(d) = \frac{4\pi}{3}d^3 \bigl[\SFTF(d+\Delta d) - \SFTF(d)\bigr]
\]
gives the number of objects within $d$ that are observable between $d$ and $d+\Delta d$.
Therefore, to first order in $\Delta d$,
\begin{equation}
\frac{\dd \ln \SFTF}{\dd d} \Delta d = -\frac{F(d)}{T(d)} \; ,
\end{equation}
where both $F$ and $T$ can be obtained directly from the data.
For the special case $S_\eta =1$ for $\eta >\eta_{l0} =const$ then 
$\SFTF(d)=const$ for $\eta_l(d) <\eta_{l0}$ leading to unbiased fTF distance modulus for objects with 
small $d$ satisfying $\muFTF \le -a \eta_{l0}+m-b$.

\section{Velocity-gravity comparisons}

\label{sec:vgcomp}
Several of these comparisons use observed $\vobs$ in peculiar–velocity catalogs to be contrasted with independent predictions of the peculiar–velocity field inferred from galaxy redshift surveys \citep{Davis2011,carrick15,Lilow2024}. In these comparisons, the predicted field is interpolated at the positions of galaxies in the peculiar–velocity catalog, and the interpolated predictions are compared with the observed velocities. 
The purpose of such comparisons is to constrain the growth rate of cosmological perturbations, which sets the overall amplitude of the predicted velocity field.

The question arises again of where to place the galaxies in the peculiar–velocity catalog. As we have seen, any field interpolated in a distance other than the true one is biased to some extent.

A common and practical approach is to use the redshifts as proxies for the true distances and place the galaxies at their redshift coordinates $s$. Since, to first order, $V(s)=V(r+V)\approx V(r)$, this approximation is valid on large scales, away from regions dominated by incoherent motions. Consider therefore the field
\[
\conm{\vobs=s-d}{s}=s-\conm{d}{s}\; .
\]
Our goal is to understand the behavior of $\conm{d}{s}$ and to what extent it approximates the true distance.

From the law of total expectation,
\[
\conm{d}{s}=\int \conm{d}{r}\,P(r|s)\,{\rm d}r\; .
\]
For an unbiased distance estimator (as in the iTF, or in the FTF after bias correction), $\conm{d}{r}=r$, hence
\[
\conm{d}{s}=\int r\,P(r|s)\,{\rm d}r=\conm{r}{s}\; .
\]
However, $\conm{r}{s}$ is not the same as the real–space solution $r_t$ of $s=r_t+V(r_t)$. The conditional distribution
\[ P(r|s)\propto P(s|r)\,P(r)\propto
\exp\!\left[-\frac{(s-r-V(r))^{2}}{2\sigma_v^{2}}\right]\,P(r)\]
depends on the small–scale velocity dispersion $\sigma_v$ and on the prior $P(r)\propto r^{2}n(r)S(r)$. Expanding the posterior about $r_t$ gives
\[
\conm{r}{s}
= r_t
+\frac{\sigma_v^{2}}{\bigl(1+V’(r_t)\bigr)^{2}}\,\partial_r\ln P(r)\big|{r_t}
\; .
\]
 Assuming further $|V’|\ll 1$
\[
\conm{\ves}{s}
= s-\conm{d}{s}
= s-\conm{r}{s}
\approx  V(r_t)
-\frac{\sigma_v^{2}}{r_t}\,\pdv{\ln P(r)}{\ln r}\big|{r_t}
\; .
\]

Therefore, when galaxies are placed at their redshift coordinates, the mean observed field $\conm{\ves}{s}$ traces the true velocity field to leading order, with a small systematic correction governed by $\sigma_v^{2}/r_t$ and the gradient of $P(r)$. Since this shift is uncorrelated with the 
the velocity field, it does not affect the inference of the growth factor from the observed versus predicted velocities.

\section{THE DISTRIBUTION OF $\hat{r}$ GIVEN $r$}
\label{sec:rhatr}
We derive the distribution of posterior samples $\hat{r}$ given the true distance $r$. 
The posterior samples are drawn from $P(\hat{r}|d)$ for observed distances $d$ that are 
themselves drawn from $P(d|r)$. The distribution of $\hat{r}$ given $r$ is therefore

\begin{equation}
P(\hat{r}|r) = \int \dd d \, P(\hat{r}|d) \, P(d|r) \,.
\label{eq:prhat_r_integral}
\end{equation}

The likelihood of observed distance given true distance is
\begin{equation}
P(d|r) \propto d^{-1} \mathcal{G}_{\ln}(r/d, \sigma_{\ln}) \,,
\end{equation}
and by Bayes' theorem the posterior is
\begin{equation}
P(\hat{r}|d) \propto P(d|\hat{r}) P(\hat{r}) \,,
\end{equation}
with prior $P(\hat{r}) \propto \hat{r}^2 n(\hat{r})$ and 
$P(d|\hat{r}) \propto d^{-1} \mathcal{G}_{\ln}(\hat{r}/d, \sigma_{\ln})$.

Substituting into eq.~(\ref{eq:prhat_r_integral}) and transforming to logarithmic 
variables $x = \ln d$, $\rho = \ln r$, $\hat{\rho} = \ln \hat{r}$, we obtain
\begin{equation}
P(\hat{r}|r) \propto \hat{r}^2 n(\hat{r}) \int \dd x \, 
\exp\left(-\frac{(\rho - x)^2}{2\sigma_{\ln}^2}\right) 
\exp\left(-\frac{(\hat{\rho} - x)^2}{2\sigma_{\ln}^2}\right) .
\end{equation}
The product of the two Gaussians in $x$ yields
\begin{equation}
\exp\left(-\frac{(\rho - x)^2 + (\hat{\rho} - x)^2}{2\sigma_{\ln}^2}\right) 
= \exp\left(-\frac{(x - \bar{x})^2}{\sigma_{\ln}^2} 
- \frac{(\rho - \hat{\rho})^2}{4\sigma_{\ln}^2}\right) ,
\end{equation}
where $\bar{x} = (\rho + \hat{\rho})/2$. Performing the Gaussian integral over $x$ gives 
the final result:
\begin{equation}
P(\hat{r}|r) = \frac{\hat{r}^2 n(\hat{r}) \, \mathcal{G}_{\ln}(r/\hat{r}, \sqrt{2}\sigma_{\ln})}
{\int \dd \hat{r}' \, \hat{r}'^2 n(\hat{r}') \, \mathcal{G}_{\ln}(r/\hat{r}', \sqrt{2}\sigma_{\ln})} \,.
\label{eq:prhat_r_final}
\end{equation}

This result has two important implications. First, the effective scatter in 
$P(\hat{r}|r)$ is $\sqrt{2}\sigma_{\ln}$, reflecting the convolution of the two 
lognormal distributions in eq.~(\ref{eq:prhat_r_integral}). Second, the weighting 
by $\hat{r}^2 n(\hat{r})$ ensures that $\langle \hat{r} | r \rangle \neq r$ for any 
density field that is not proportional to $r^{-2}$. For uniform density 
$n(\hat{r}) = \text{const}$,
\begin{equation}
\langle \ln \hat{r} | r \rangle = \ln r + 6\sigma_{\ln}^2 \,,
\end{equation}
confirming that posterior samples are systematically biased to larger distances.

\section{Assessment of Gibbs Sampling for Joint Distance-Velocity Inference}
\label{sec:graziani}

\citet{Graziani_Hoffman} proposed a Bayesian approach to jointly infer galaxy distances and the velocity field using Gibbs sampling. Following their method and that of \citet{SW95}, we write the likelihood of observing distance $d$ and redshift $s$ given true distance $r$ and velocity field $V$ as
\begin{equation}
\label{eq:PdsrV}
\begin{split}
P(d,s\mid r,V) \propto 
\prod_{i} &\exp\!\left[-\frac{\ln^{2}(d_i/r_i)}{2\sigln^{2}}\right] \\
&\times \exp\!\left[-\frac{(s_i-r_i-V(r_i))^{2}}{2\sigma_{s}^{2}}\right] P(V),
\end{split}
\end{equation}
where the product extends over galaxies in the distance-indicator catalog and $P(V)$ denotes a prior on the velocity field, assumed to be a multivariate normal distribution with covariance fixed by the cosmological model.

Note the difference in notation from \citet{Graziani_Hoffman}: in their convention $d_i$ denotes the true luminosity distance, whereas here it represents the \emph{observed} distance. Since we work in the low-redshift limit, we approximate the luminosity distance by $r_i$. Thus, the PDF term in their equation~(13), which involves the observed distance modulus and the luminosity distance, corresponds to our term in $P[\ln(r/d)]$ in Eq.~(\ref{eq:PdsrV}).

\subsection{The Gibbs Sampling Procedure}

Using a PDF of the form given in Eq.~(\ref{eq:PdsrV}), \citet{Graziani_Hoffman} perform Gibbs sampling over $\{r_i,V_i\}$. Their procedure can be summarized as follows:
\begin{enumerate}
\item For a given set of $r_i$, compute $V_i^{r}=s_i-r_i$ and use it as input to the Hoffman--Ribak algorithm \citep{hr91} to obtain a constrained realization of the velocity field on a grid, $V_\mathrm{grid}$.

\item Given $V_\mathrm{grid}$, sample each $r_i$ from the posterior $P(r_i|d_i, s_i, V_\mathrm{grid})$, where evaluating this posterior at a particular $r_i$ requires interpolating $V_\mathrm{grid}$ to that position.
\end{enumerate}
The process is iterated to generate joint samples from the posterior distribution $P(\{r_i,V_i\}|\{d_i, s_i\})$.

\subsection{Behavior in the Large Distance-Error Regime}

To examine how this approach performs when distance errors are substantial, we consider a single observed galaxy with measured redshift $s$ and distance $d$. For simplicity, we restrict the analysis to one line of sight and approximate $\ln(r/d) \simeq (r-d)/d$, thereby replacing the lognormal distribution with a Gaussian of width $\sigma_d = d\,\sigln$. We assume a velocity covariance of exponential form,
\begin{equation}
\langle V(r_1) V(r_2)\rangle  = C_0\,\exp(-|r_1-r_2|/r_s)\, ,
\end{equation}
and implement the Gibbs sampling procedure with the following parameters: observed distance $d = 90$~Mpc, redshift $s = 100$~Mpc, distance error $\sigma_d = 13.5$~Mpc, velocity dispersion $\sigma_v = 2$~Mpc (equivalent to 200~km\,s$^{-1}$), velocity covariance amplitude $C_0 = 4$~Mpc$^2$, and coherence scale $r_s = 10$~Mpc.

The results are shown in Figure~\ref{fig:rV_panels}. Despite the observed distance $d = 90$~Mpc differing significantly from the redshift $s = 100$~Mpc, the Gibbs sampler converges to a posterior mean distance of $\langle r \rangle = 99.6$~Mpc---essentially the redshift value. The posterior distribution (panel~a) is tightly concentrated near $s$, with negligible weight at the observed distance $d$. Similarly, the velocity samples (panel~b) cluster near zero, consistent with $V \approx s - r \approx 0$ when $r \approx s$. The joint posterior (panel~c) shows strong correlation between $r$ and $V$, tracing the constraint $s = r + V$.

This behavior reflects a regime limitation: when distance errors are large, the strong prior from the redshift measurement, combined with the velocity field prior, dominates over the distance information from the TF relation. The algorithm preferentially places galaxies near their redshift coordinates, where peculiar velocities are minimized.

\subsection{Analytical Understanding}

To understand the origin of this behavior, we examine the joint posterior analytically. For simplicity, we adopt a univariate normal prior $P(V) \propto \exp(-V^2/(2\sigma_V^2))$ and write
\begin{equation}
P(d,s\mid r,V) \propto 
\exp\!\left[-\frac{(r-d)^{2}}{2\sigma_{d}^{2}}\right]
\exp\!\left[-\frac{(s-r-V)^{2}}{2\sigma_{v}^{2}}
-\frac{V^{2}}{2\sigma_{V}^{2}}\right].
\end{equation}

Maximizing with respect to $r$ and $V$ gives
\begin{equation}
r_{\max} = \frac{d\,(\sigma_{v}^{2}+\sigma_{V}^{2}) + s\,\sigma_{d}^{2}}
{(\sigma_{v}^{2}+\sigma_{V}^{2}) + \sigma_{d}^{2}}
\end{equation}
and 
\begin{equation}
V_{\max} = \frac{\sigma_{V}^{2}}{\sigma_{v}^{2}+\sigma_{V}^{2}+\sigma_d^2}\,(s - d)\, .
\end{equation}

\begin{figure*}[ht!]
\gridline{
\fig{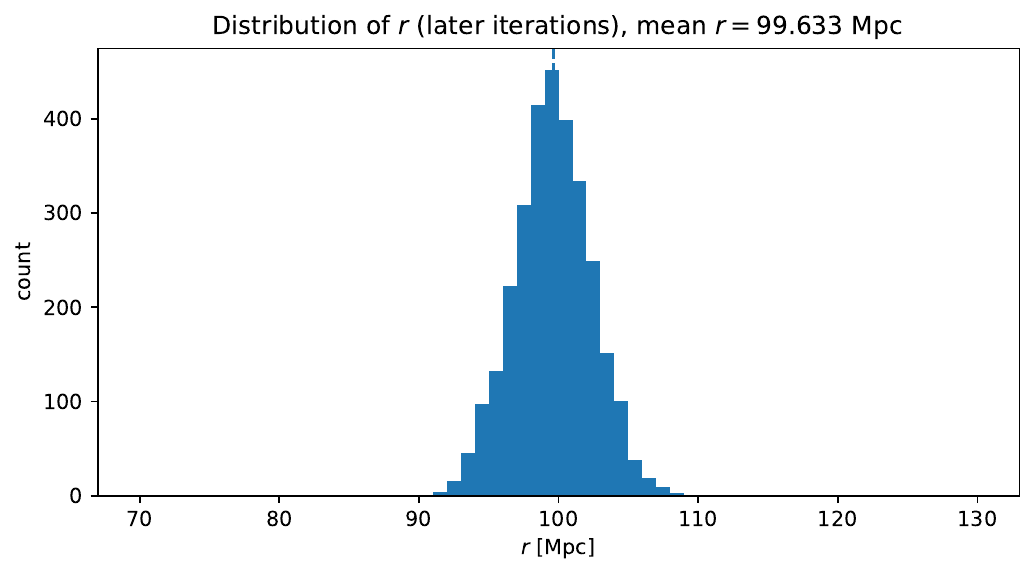}{0.45\textwidth}{(a)}
\fig{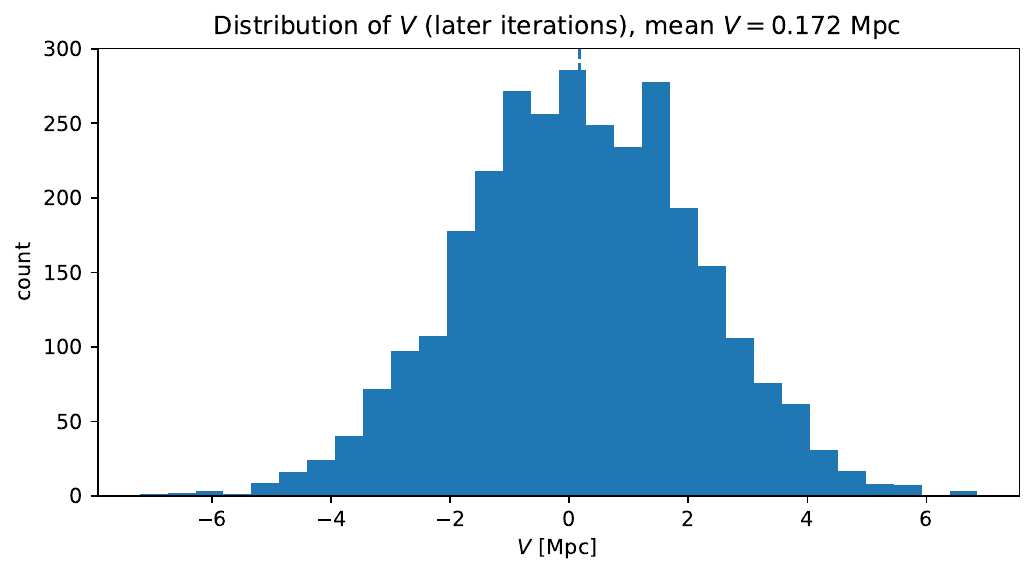}{0.45\textwidth}{(b)}
}
\gridline{
\fig{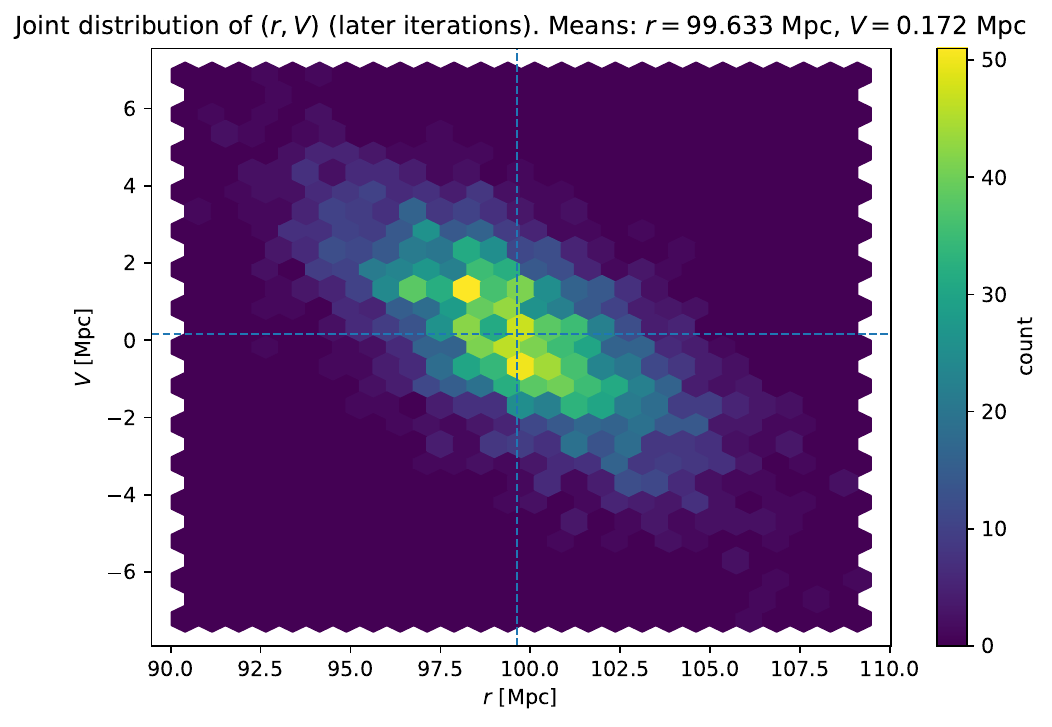}{0.45\textwidth}{(c)}
\fig{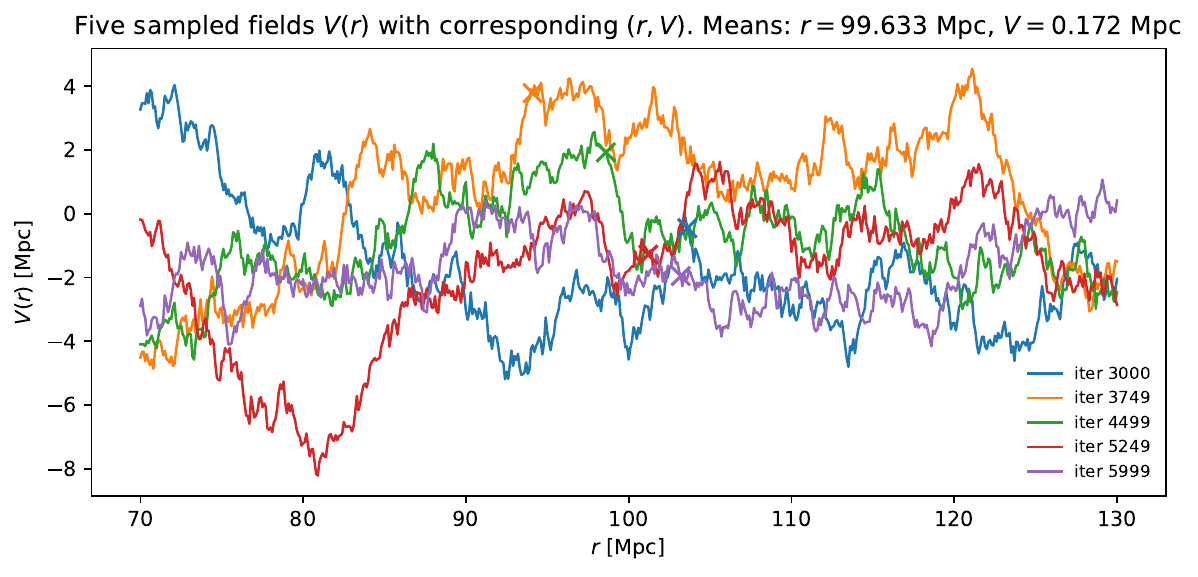}{0.45\textwidth}{(d)}
}
\caption{
Gibbs sampling results for a single galaxy with observed distance $d=90$~Mpc and redshift $s=100$~Mpc.
(a)~Posterior distribution of sampled distances $r$; dashed line marks the mean. 
(b)~Posterior distribution of sampled velocities $V$; dashed line marks the mean.
(c)~Joint posterior distribution of $(r,V)$ shown as hexbin density; dashed lines indicate the means.
(d)~Five representative velocity field realizations $V(r)$ from late Gibbs iterations, with corresponding sampled points $(r,V)$ marked by crosses.
Parameters: velocity covariance amplitude $C_0=4$~Mpc$^2$, exponential coherence scale $r_s=10$~Mpc, distance error $\sigma_d=13.5$~Mpc, and nonlinear velocity dispersion $\sigma_v=2$~Mpc ($200$~km\,s$^{-1}$).
\label{fig:rV_panels}}
\end{figure*}

At large distances where $\sigma_d \gg \sigma_V, \sigma_v$, we have $r_{\max}\approx s$ and hence $V_{\max}\approx 0$. This explains the behavior observed in Figure~\ref{fig:rV_panels}: when distance errors dominate, the posterior is driven primarily by the redshift constraint, and the distance measurement contributes little additional information.

We emphasize that this limitation is not a deficiency in the Gibbs sampling implementation itself, but rather reflects the fundamental information content available when distance errors dominate the error budget. The method may perform well in regimes where distance uncertainties are comparable to or smaller than the velocity field scales. However, for typical distance-indicator surveys at cosmological distances---where $\sigma_d$ grows with distance and can substantially exceed both $\sigma_v$ and $\sigma_V$---the posterior becomes increasingly dominated by the redshift prior, and the inferred distances converge toward redshift-space coordinates. In this regime, the results approach those of the simpler redshift-space reconstruction discussed in \S\ref{sec:velocity_redshift}, suggesting that alternative approaches may be more effective for extracting velocity field information from such data.

\bibliography{MB}{}

\begin{thebibliography}{}
\expandafter\ifx\csname natexlab\endcsname\relax\def\natexlab#1{#1}\fi
\providecommand{\url}[1]{\href{#1}{#1}}
\providecommand{\dodoi}[1]{doi:~\href{http://doi.org/#1}{\nolinkurl{#1}}}
\providecommand{\doeprint}[1]{\href{http://ascl.net/#1}{\nolinkurl{http://ascl.net/#1}}}
\providecommand{\doarXiv}[1]{\href{https://arxiv.org/abs/#1}{\nolinkurl{https://arxiv.org/abs/#1}}}

\bibitem[{Aaronson {et~al.}(1982)Aaronson, Huchra, Mould, Tully, Fisher, van
  Woerden, Goss, Chamaraux, Mebold, Siegman, Berriman, \&
  Persson}]{Aaronson1982}
Aaronson, M., Huchra, J., Mould, J.~R., {et~al.} 1982, ApJS, 50, 241,
  \dodoi{10.1086/190827}

\bibitem[{Branchini {et~al.}(2012)Branchini, Davis, \& Nusser}]{BND12}
Branchini, E., Davis, M., \& Nusser, A. 2012, MNRAS, 424, 472,
  \dodoi{10.1111/J.1365-2966.2012.21210.X}

\bibitem[{Carrick {et~al.}(2015)Carrick, Turnbull, Lavaux, \&
  Hudson}]{carrick15}
Carrick, J., Turnbull, S.~J., Lavaux, G., \& Hudson, M.~J. 2015, MNRAS, 450,
  317, \dodoi{10.1093/mnras/stv547}

\bibitem[{Davis {et~al.}(1982)Davis, Huchra, Latham, \& Tonry}]{dav82}
Davis, M., Huchra, J., Latham, D.~W., \& Tonry, J. 1982, {\textbackslash}apj,
  253, 423, \dodoi{10.1086/159646}

\bibitem[{Davis {et~al.}(2011)Davis, Nusser, Masters, Springob, Huchra, \&
  Lemson}]{Davis2011}
Davis, M., Nusser, A., Masters, K.~L., {et~al.} 2011, MNRAS, 413, 2906,
  \dodoi{10.1111/j.1365-2966.2011.18362.x}

\bibitem[{Djorgovski \& Davis(1987)}]{Fundamental_Plan87}
Djorgovski, S., \& Davis, M. 1987, ApJ, 313, 59, \dodoi{10.1086/164948}

\bibitem[{Feast(1972)}]{Feast1972}
Feast, M.~W. 1972, Vistas in Astronomy, 13, 207,
  \dodoi{10.1016/0083-6656(72)90013-X}

\bibitem[{Graziani {et~al.}(2019)Graziani, Courtois, Lavaux, Hoffman, Tully,
  Copin, \& Pomar{\`{e}}de}]{Graziani_Hoffman}
Graziani, R., Courtois, H.~M., Lavaux, G., {et~al.} 2019, MNRAS, 488, 5438,
  \dodoi{10.1093/mnras/stz078}

\bibitem[{Hoffman {et~al.}(2015)Hoffman, Courtois, \& Tully}]{Hoffman15}
Hoffman, Y., Courtois, H.~M., \& Tully, R.~B. 2015, MNRAS, 449, 4494,
  \dodoi{10.1093/mnras/stv615}

\bibitem[{Hoffman {et~al.}(2021)Hoffman, Nusser, Valade, Libeskind, \&
  Tully}]{Hoffman2021}
Hoffman, Y., Nusser, A., Valade, A., Libeskind, N.~I., \& Tully, R.~B. 2021,
  MNRAS, 505, 3380, \dodoi{10.1093/mnras/stab1457}

\bibitem[{Hoffman \& Ribak(1991)}]{hr91}
Hoffman, Y., \& Ribak, E. 1991, ApJL, 380, L5, \dodoi{10.1086/186160}

\bibitem[{Landy \& Szalay(1992)}]{Landy1992}
Landy, S.~D., \& Szalay, A.~S. 1992, ApJ, 391, 494, \dodoi{10.1086/171365}

\bibitem[{Lilow {et~al.}(2024)Lilow, Ganeshaiah~Veena, Nusser, Lilow,
  Ganeshaiah~Veena, \& Nusser}]{Lilow2024}
Lilow, R., Ganeshaiah~Veena, P., Nusser, A., {et~al.} 2024, arXiv,
  arXiv:2404.02278, \dodoi{10.48550/ARXIV.2404.02278}

\bibitem[{Lilow {et~al.}(2021)Lilow, Nusser, Lilow, \&
  Nusser}]{LilowNusser2021}
Lilow, R., Nusser, A., Lilow, R., \& Nusser, A. 2021, MNRAS, 507, 1557,
  \dodoi{10.1093/MNRAS/STAB2009}

\bibitem[{Lynden-Bell {et~al.}(1988)Lynden-Bell, Faber, Burstein, Davies,
  Dressler, Terlevich, \& Wegner}]{Lyn88}
Lynden-Bell, D., Faber, S.~M., Burstein, D., {et~al.} 1988, ApJ, 326, 19,
  \dodoi{10.1086/166066}

\bibitem[{Nusser {et~al.}(2011)Nusser, Branchini, \& Davis}]{NBDL}
Nusser, A., Branchini, E., \& Davis, M. 2011, ApJ, 735, 77,
  \dodoi{10.1088/0004-637X/735/2/77}

\bibitem[{Nusser \& Davis(1995)}]{Nusser1995}
Nusser, A., \& Davis, M. 1995, MNRAS, 276, 1391

\bibitem[{Nusser {et~al.}(2020)Nusser, Yepes, \& Branchini}]{Nusser2020}
Nusser, A., Yepes, G., \& Branchini, E. 2020, ApJ, 905, 47,
  \dodoi{10.3847/1538-4357/abc42f}

\bibitem[{Peebles(1980)}]{Peeb80}
Peebles, P. J.~E. 1980, {The large-scale structure of the universe} (Princeton
  University Press, NJ).
\newblock \url{http://adsabs.harvard.edu/abs/1980lssu.book.....P}

\bibitem[{Saulder {et~al.}(2013)Saulder, Mieske, Zeilinger, \&
  Chilingarian}]{Saulder_Fundamental}
Saulder, C., Mieske, S., Zeilinger, W.~W., \& Chilingarian, I. 2013, AA, 557,
  A21, \dodoi{10.1051/0004-6361/201321466}

\bibitem[{Strauss \& Willick(1995)}]{SW95}
Strauss, M.~A., \& Willick, J.~A. 1995, Physics Reports, 261, 271,
  \dodoi{10.1016/0370-1573(95)00013-7}

\bibitem[{Tully \& Fisher(1977)}]{TF77}
Tully, R.~B., \& Fisher, J.~R. 1977, {\textbackslash}aap, 54, 661

\bibitem[{Tully {et~al.}(2023)Tully, Kourkchi, Courtois, Anand, Blakeslee,
  Brout, Jaeger, Dupuy, Guinet, Howlett, Jensen, Pomar{\`{e}}de, Rizzi, Rubin,
  Said, Scolnic, \& Stahl}]{CF4}
Tully, R.~B., Kourkchi, E., Courtois, H.~M., {et~al.} 2023, ApJ, 944, 94,
  \dodoi{10.3847/1538-4357/ac94d8}

\bibitem[{Veena {et~al.}(2023)Veena, Lilow, \& Nusser}]{Punya2023}
Veena, P.~G., Lilow, R., \& Nusser, A. 2023, MNRAS, 522, 5291,
  \dodoi{10.1093/mnras/stad1222}

\bibitem[{Watkins \& Feldman(2015)}]{Watkins2015}
Watkins, R., \& Feldman, H.~A. 2015, MNRAS, 450, 1868,
  \dodoi{10.1093/mnras/stv651}

\bibitem[{Zaroubi {et~al.}(1999)Zaroubi, Hoffman, \&
  Dekel}]{1999ApJ...520..413Z}
Zaroubi, S., Hoffman, Y., \& Dekel, A. 1999, {\textbackslash}apj, 520, 413,
  \dodoi{10.1086/307473}

\end{thebibliography}
\bibliographystyle{aasjournal}



\end{document}